\documentclass[12pt]{article}
\usepackage{newinutile}
\usepackage{amsmath,amssymb}
\usepackage{epsfig,graphics,color,calc}
\usepackage{cite}
\usepackage{rotating}

\newtheorem{theorem}{Theorem}

\newtheorem{lemma}[theorem]{Lemma}

\newtheorem{corollary}[theorem]{Corollary}

\newtheorem{definition}[theorem]{Definition}

\newtheorem{condition}[theorem]{Condition}

\newcommand{\newatop}[2]{\genfrac{}{}{0pt}{}{#1}{#2}}
\newcommand{\qed}{$\phantom.$\hfill $\Box$\bigskip}
\newcommand{\rr}[1]{{\normalfont\textrm{#1}}}
\newcommand{\cc}[1]{{\mathcal{#1}}}
\newcommand{\bb}[1]{{\mathbb{#1}}}

\newcommand{\puno}{{\normalfont\textbf{u}}}
\newcommand{\muno}{{\normalfont\textbf{d}}}
\newcommand{\zero}{{\normalfont\textbf{0}}}
\newcommand{\pcri}{{\cc{P}_\mathrm{c}}}
\newcommand{\qcri}{{\cc{Q}_\mathrm{c}}}
\newcommand{\vbim}{\cc{X}_\muno}
\newcommand{\vbiminore}{\cc{X}_{\muno,<}}
\newcommand{\vbimaggiore}{\cc{X}_{\muno,>}}
\newcommand{\vbimzero}{\cc{X}_{\muno,\zero}}
\newcommand{\vbizero}{\cc{X}_\zero}

\newcommand{\definisco}[1]{{\em{#1}}}

\newlength{\pecettawidth}
\begin{document}
\title{Relaxation Height in Energy Landscapes: an Application \\ 
       \vskip 0.3 cm 
       to Multiple Metastable States}

\author{Emilio N.M.\ Cirillo}
\affiliation{Dipartimento di Scienze di Base e Applicate per 
             l'Ingegneria, Sapienza Universit\`a di Roma, 
             via A.\ Scarpa 16, I--00161, Roma, Italy.}
\email{emilio.cirillo@uniroma1.it}
\thanks{This work has been partially done at Eurandom during the 
Stochastic Activity Month, February 2012. 
ENMC wants to express his thanks to the organizers, R.\ Fernandez, 
R.\ van der Hofstad, and M.\ Heydenreich, for the invitation and to 
Eurandom for the kind hospitality. FRN thanks F.\ den Hollander and 
A.\ Troiani for 
many interesting discussions related to the topic of the paper.}

\author{Francesca R.\ Nardi}
\affiliation{Department of Mathematics and Computer Science,
             Eindhoven University of Technology,
             P.O.\ Box 513, 5600 MB Eindhoven, The Netherlands.}
\affiliation{Eurandom, P.O.\ Box 513, 5600 MB, Eindhoven, The Netherlands.}
\email{F.R.Nardi@tue.nl}


\begin{abstract}
The study of systems with multiple (not necessarily degenerate) 
metastable states 
presents subtle difficulties from the mathematical point of view
related to the variational problem that has to 
be solved in these cases. 
We introduce the notion of relaxation height in a general energy 
landscape and we prove
sufficient conditions which are valid even in presence of multiple 
metastable states. 
We show how these results can be used to approach the problem of multiple 
metastable states via the use of the modern theories of 
metastability. 
We finally apply these general results to the 
Blume--Capel model for a particular choice of the parameters 
ensuring the existence of two multiple, and not degenerate in energy, 
metastable states.
\end{abstract}



\preprint{Appunti: \today}


\maketitle

\section{Introduction}
\label{s:int}
\par\noindent
In many applicative problems one has to consider a stochastic 
system evolving in a finite not empty state space 
driven by an energy function. The details of the dynamics depend
on the system that one has to model, but in most cases the motion 
is driven by the energy landscape \cite{W}. 
The typical trajectories are those decreasing the energy;
on the other hand  
with small probability the system can perform 
jumps against the energy drift. 

The smallness of the probability associated to moves against 
the energy drift is controlled by some cost function. This cost  
is often given by the difference of energy 
between the two states involved in the move. This is 
the situation that one has to face when studying the evolution 
of Glauber dynamics associated with Statistical Mechanics lattice 
models \cite{Glauber}.
Depending on the model, different cost functions can be introduced.
An example, which deserved a lot of attention in the recent 
literature, is that of reversible \cite{GJH}
Probabilistic Cellular Automata 
\cite{BCLS,CN,CNP,NS,CNS01,CNS02}.

The stochastic dynamics is often controlled by a parameter, 
say the {\em temperature}, whose value tunes the amount 
of randomness in the motion by increasing or 
decreasing the probability of the moves against the drift. 
When the temperature is very low the system freezes and tends to stick to the 
energy drift dynamics ending up to be trapped 
in the minima of the energy. When the temperature is very high, on the other 
hand, jumps against the drift are highly probable and the system 
moves almost freely in the state space. 

It is then clear that a huge amount of information about the 
low temperature dynamics is obtained when one knows 
(i) the 
structure of the absolute minima of the energy 
(ground states) 
and 
(ii) the maximal barrier that has to be overcome 
to reach the set of the ground states starting from any possible state, 
namely, the \definisco{relaxation height}.
The severity of problem (i) strongly depends 
on the particular expression of the energy function, 
but in most applications it is not a particularly difficult task. 
Problem (ii), on the other hand, is always a very laborious
(and sometimes hard) one
since the solution of several variational problems is involved. 

The relaxation height
is of basic importance in the study 
of this kind of dynamics.
For instance, 
when dealing with the description of the metastable 
behavior of Statistical Mechanics systems, the relaxation 
height is the quantity controlling the typical time 
the system waits before nucleating the stable state starting from
the metastable one \cite{MNOS,BEGK}.
The problem of computing the relaxation height in specific model 
is often very hard, in particular when there are several 
states connected to the ground states via 
cost barriers equal to the relaxation height. 
This situation is found, for instance, in models 
with more than one metastable state \cite{CN}.

In this paper we shall discuss some generic properties of the relaxation 
height and, in particular, we shall setup a quite general strategy 
for its computation. Subsequently we shall apply the 
theory to the very interesting case of the Blume--Capel 
model \cite{B,C,BEG,CO} for a choice of the parameter 
ensuring the existence of two not degenerate in energy 
metastable states. Due to the presence 
of multiple metastable states this case is particularly 
interesting and, a priori, very difficult to be treated.

The Blume--Capel model has been originally introduced in 
\cite{B,C} to study some particular magnetic systems. 
The model was then generalized to the so called 
Blume--Emery--Griffiths model \cite{BEG} in order to describe the 
$\lambda$ transition in He$^3$--He$^4$ mixtures.
Those models have been widely studied in the literature and 
many applications have been considered. Their particular interest 
is due to fact that in these three state spin systems 
on lattice different energies are associated with different 
interfaces. This is not the case in other multi--state spin systems, 
such as the well known Potts model, where a ferromagnetic coupling 
favors the presence of neighboring homologous spins and 
the same energy cost is assigned to any pair of differing
neighboring spins. 

This fact is very peculiar of the Blume--Capel model and 
gives rise to very interesting phenomena when the metastable 
behavior of the system is considered. 
Indeed it is seen that even if one of the three states is not 
favored from the energetic point of view, it can (and indeed does) 
behave as a bearing between the other two phases if 
the interfaces between this particular state and the other two ones are 
energetically less expensive (see, for instance, 
the proof of item~\ref{i:mdep-bc01-02} in Lemma~\ref{t:mdep-bc01}
where this effect appears in all its importance).

In \cite{CO} the authors 
studied metastability in the Blume--Capel model in the so called 
Wentzell--Friedlin regime, that is they considered finite 
volume, finite positive magnetic field, and let the temperature 
tend to zero. 
In that paper a wide region of the space of parameters of the model was 
analyzed. More precisely, denoted by $\lambda$ the chemical potential
(see equation (\ref{ham-bc}) below) and by $h$ the magnetic field, the region 
$h>\lambda>0$ was studied in detail. 
It was proven that in that region 
the Hamiltonian of the system has a single ground state $\puno$, namely, 
the configuration with all the spins equal to plus one, 
and that the system admits a single 
metastable state, that is the configuration $\muno$ with all the spins 
equal to minus one. 
It was also proven that the configuration $\zero$, with all 
the spins equal to zero, plays an important role. 
For $2\lambda>h>\lambda$ the transition from the metastable state 
$\muno$ to the stable one $\puno$ is direct, although the plus droplet 
growing inside the minus phase is ``protected" by a thin frame of 
zeros. 
For $h>2\lambda$, on the other hand, during the transition 
the system visits the intermediate (not metastable) state $\zero$.
This kind of phenomena is not expected to be observed 
in multi--state spin systems such as the Potts model where 
all the interfaces pay the same energy cost. 

In \cite{MO} the study of the metastable behavior of the Blume--Capel model 
was addressed in infinite volume in 
the regime characterized by finite $\lambda$ and $h$ and 
temperature tending to zero. In that paper the authors 
studied the regions $0<-\lambda<h$ and $0<\lambda<h$ of the 
space of parameters,
see \cite[Theorem~1]{MO}, and proved that, starting from $\muno$,  
in the first one the state $\zero$ is observed before the transition 
to $\puno$, while in the second one it is not. 

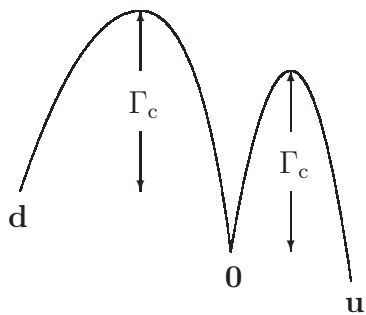
\begin{figure}[t]
 \begin{picture}(100,130)(-50,25)
 \thinlines
 \setlength{\unitlength}{0.08cm}
 \qbezier(50,30)(60,60)(70,60)
 \qbezier(70,60)(80,60)(85,20)
 \put(70,50){\vector(0,1){10}}
 \put(70,40){\vector(0,-1){10}}
 \put(68,43){${\Gamma_\rr{c}}$}
 \put(48,24){${\muno}$}
 \put(84,14){${\zero}$}
 \qbezier(85,20)(90,50)(95,50)
 \qbezier(95,50)(100,50)(105,15)
 \put(95,40){\vector(0,1){10}}
 \put(95,30){\vector(0,-1){10}}
 \put(93,33){${\Gamma_\rr{c}}$}
 \put(104,10){${\puno}$}
 \end{picture}
 \caption{Schematic description of the energy landscape for the 
          Blume--Capel model (\ref{ham-bc}) at $\lambda=0$ and $h>0$.
          The quantity $\Gamma_\rr{c}$ represents the energy barrier that 
          must be overcome to exit the metastable states. In the 
          picture it is remarked that the two metastable states $\muno$ 
          and $\zero$ are not 
          degenerate in energy.}
 \label{f:fig01}
\end{figure}

This result, looked at from the finite volume point of view, 
suggests that at negative $\lambda$ the state $\zero$ is metastable 
while at positive $\lambda$ it is not. 
The first of these two remarks was discussed, 
on heuristics grounds, in \cite{CNS03}, while the second 
was indeed proven in \cite{CO}.
We can then conjecture that in the peculiar case $\lambda=0$ and 
$h>0$ both $\muno$ and $\zero$ are metastable. In other words 
we expect that for this choice of the parameters the Blume--Capel 
model shows the very interesting phenomenon of multiple 
not degenerate in energy metastable states (see figure~\ref{f:fig01}). 

Situations of this kind have been already studied in the case 
of Probabilistic Cellular Automata \cite{CN}, but there an old
approach \cite{OS} to
metastability had been used. In that paper 
the model dependent study had to be very detailed and, hence, 
very difficult.
In this paper we show that by using the more recent approach to metastability,
that will be shortly reviewed at the beginning of Section~\ref{s:bcmodel}
and the general result on relaxation height in 
Theorem~\ref{t:schemamnos} below,
the problem can be solved relying on few model dependent properties. 

The paper is organized as follows. The general results are discussed
in Section~\ref{s:genris}. Their application to the 
problem of metastability in Statistical Mechanics systems 
is discussed in Section~\ref{s:metlat}. In particular, 
their application to the study of the metastable behavior of 
the Blume--Capel model is discussed in Section~\ref{s:bcmodel}. 
Finally we prove the theorems and the lemmas in 
Sections~\ref{s:dimo} and \ref{s:dimo-meta}.

\section{Maximal stability level}
\label{s:genris}
\par\noindent
In this section we shall discuss some general results related 
to energy landscapes. Under very general assumptions on the structure of 
the energy landscape we shall introduce the notion of 
{\em maximal stability level} and prove some results 
yielding a handy recipe for its computation.

\subsection{Energy landscape}
\label{s:rel}
\par\noindent
An \definisco{energy landscape} is a quaternion
$(X,Q,H,\Delta)$ where 
the finite not empty set $X$,
$Q\subset X\times X$,
$H:X\to\bb{R}$, and
$\Delta:Q\to\bb{R}_+$ 
are respectively called 
\definisco{state space},
\definisco{connectivity relation},
\definisco{energy},
and
\definisco{cost function}, 
and the relation $Q$ on $X$ is 
such that 
for any $x,y\in X$ there exist an integer $n\ge2$ and $x_1,\dots,x_n\in X$ 
such that $x_1=x$, $x_n=y$, and $(x_i,x_{i+1})\in Q$ for any 
$i=1,\dots,n-1$. 

An energy landscape $(X,Q,H,\Delta)$ is called \definisco{reversible}
if and only if 
the connectivity relation $Q$ is symmetric
and 
\begin{equation}
\label{rev}
H(x)+\Delta(x,y)=\Delta(y,x)+H(y)
\end{equation}
for all $(x,y)\in Q$.

\subsection{Definitions}
\label{s:def}
\par\noindent
Consider a reversible energy landscape $(X,Q,H,\Delta)$.
Given $Y\subset X$ such that $H(y)=H(y')$ for any $y,y'\in Y$, 
we shall denote by $H(Y)$ the energy of the states in $Y$. 
For any $Y\subset X$ we shall denote by $F(Y)$ the set of the minima 
of the energy inside $Y$, that is to say $y\in F(Y)$ if and only 
if 
$H(y')\ge H(y)$ for any $y'\in Y$.
We let $X_\rr{s}:=F(X)$ be the set of \definisco{ground states} of $H$, 
namely, the set of the absolute minima of the energy.

For any positive integer $n$, $\omega\in X^n$ 
such that $(\omega_i,\omega_{i+1})\in Q$ for all $i=1,\dots,n-1$ 
is called a \definisco{path} joining $\omega_0$ to $\omega_n$; 
we also say that $n$ is the length of the path.
For any path $\omega$ of length $n$, we let 
\begin{equation}
\label{height}
\Phi_\omega:=\max_{i=1,\dots,n-1}H(\omega_i)+\Delta(\omega_i,\omega_{i+1})
\end{equation}
be the \definisco{height} of the path\footnote{Since the energy landscape 
is reversible, the energy of the state $\omega_n$ is implicitly taken
into account in (\ref{height}), indeed (\ref{rev}) implies 
$H(\omega_n)\le\Delta(\omega_{n-1},\omega_n)+H(\omega_{n-1})$.}. 
For any $y,z\in X$ we denote by $\Omega(y,z)$ 
the set of the paths joining $y$ to $z$ and 
define the \definisco{communication height} 
between $y$ and $z$ as 
\begin{equation}
\label{communication}
\Phi(y,z)
:=
\min_{\omega\in\Omega(y,z)}
\Phi_\omega
\end{equation}
From (\ref{rev}), (\ref{height}), and (\ref{communication}) 
it follows immediately that 
\begin{equation}
\label{rev02}
\Phi(y,z)=\Phi(z,y)
\end{equation}
for all $y,z\in X$. 
For any $Y,Z\subset X$ we let 
\begin{equation}
\label{communication-set}
\Phi(Y,Z)
:=
\min_{\omega\in\Omega(Y,Z)}\Phi_\omega
=
\min_{y\in Y,z\in Z}\Phi(y,z)
\end{equation}
where we have used the notation
$\Omega(Y,Z)$ for the set of paths joining 
a state in $Y$ to a state in $Z$. 

We say that $X$ is \definisco{fully attracted} by $X_\rr{s}$ 
if and only if 
$\Phi(x,X_\rr{s})-H(x)=0$ 
for any $x\in X\setminus X_\rr{s}$
or (trivial case) $X=X_\rr{s}$.

For any $x\in X$ 
we denote by $I_x$ the set of states $y\in X$ 
such that $H(y)<H(x)$. 
Note that $I_x=\emptyset$ if $x\in X_\rr{s}$.
We then define the 
\definisco{stability level} of any $x\in X\setminus X_\rr{s}$ 
\begin{equation}
\label{stability}
V_x:=\Phi(x,I_x)-H(x)
\ge0
\end{equation}
Note that the stability level $V_x$ of $x$ is the minimal 
cost that, starting from $x$, has to be payed
in order to reach states at energy lower than $H(x)$. 
Following \cite{MNOS} we now introduce the notion of 
maximal stability level.

\begin{definition}
\label{t:maxstab}
Consider a reversible energy landscape $(X,Q,H,\Delta)$.
Assume $X\setminus X_\rr{s}\neq\emptyset$, we let 
the \definisco{maximal stability level} be
\begin{equation}
\label{gamma}
\Gamma_\rr{m}:=\sup_{x\in X\setminus X_\rr{s}}V_x
\end{equation}
We also set 
\begin{equation}
\label{metastabile}
X_\rr{m}
:=
\{x\in X\setminus X_\rr{s}:\,V_x=\Gamma\}
\end{equation}
\end{definition}
Note that, since the state space is finite, the maximal stability level 
$\Gamma_\rr{m}$ is a finite number. 
Note, also, that if $X$ is fully attracted by 
$X_\rr{s}$ and $X\setminus X_\rr{s}\neq\emptyset$, 
then $\Gamma_\rr{m}=0$ and $X_\rr{m}=X\setminus X_\rr{s}$. 

\subsection{Results}
\label{s:res}
\par\noindent
The notion of maximal stability level has been introduced by 
looking at the paths starting from any 
state of $X$ and reaching lower energy states. 
This point of view is often very useful when dealing with 
metastability problems \cite{MNOS}, indeed the maximal stability 
level controls the asymptotic of the time the system spends in the metastable
state before nucleating the stable one. 
It is worth 
noting, on the other hand, that computing the maximal stability 
level of a concrete model is a hard task. 
Indeed, one has to solve the variational problem $\Phi(x,I_x)$ 
for any $x\in X$. 

It is then clear the interest of results providing 
sufficient conditions, whose verification 
in the context of specific model is of reasonable difficulty,
ensuring that a real number is the maximal stability level.
We note that 
this question has already been debated in the pertaining literature, 
see for instance \cite[Section~4.2]{MNOS}, \cite[Theorem~2.3]{CNS01},
\cite[Lemma~1.2 and hypothesis H2]{HNT},
and \cite[Lemmas~1.16, 1.16 and 1.17]{HNT1}. 
We remark that in all the quoted 
references the authors always stated results in the case $|X_\rr{m}|=1$.

These types of results can be obtained by looking at the problem 
by a different point of view, that is 
by looking at all the paths 
connecting any state not belonging to $X_\rr{s}$ to the 
set of ground states itself. 

\begin{theorem}
\label{t:tunn}
Consider a reversible energy landscape $(X,Q,H,\Delta)$.
Assume $X\setminus X_\rr{s}\neq\emptyset$.
If the not empty set $A\subset X\setminus X_\rr{s}$ and the 
positive real number $a\in\bb{R}_+$ are such that 
\begin{enumerate}
\item\label{i:tunn01}
$\Phi(x,X_\rr{s})-H(x)=a$
for all 
$x\in A$;
\item\label{i:tunn02}
either 
$X\setminus(A\cup X_\rr{s})=\emptyset$
or
$\Phi(x,X_\rr{s})-H(x)<a$
for all $x\in X\setminus(A\cup X_\rr{s})$;
\end{enumerate}
then 
\begin{displaymath}
\Gamma_\rr{m}=a
\;\;\;\textrm{ and }\;\;\;
X_\rr{m}=A
\end{displaymath}
\end{theorem}

The above theorem is very useful in the applications, 
indeed it gives a general strategy to approach the problem 
of computing the maximal stability level in special models. 
The idea is that one has to figure out what is the set 
of configurations such that starting from them the cost 
to reach the ground states is precisely the maximal stability level. 
Once it has been proven that starting from all 
the configurations in this set
the cost to be payed is the same, then one is just left with the proof 
that starting from any other configuration the cost is strictly smaller. 
And this is not a terrific computation since only an upper 
bound to the height along the paths is needed.

It is possible to prove a necessary condition in the spirit of
the statement in Theorem~\ref{t:tunn}. 
In other words on the basis of the Definition~\ref{t:maxstab} 
we can say what is the barrier that must be overcome to visit the set 
of ground states $X_\rr{s}$ starting from states in $X_\rr{m}$.

\begin{theorem}
\label{t:tunn-necessity}
Consider a reversible energy landscape $(X,Q,H,\Delta)$.
Assume $X\setminus X_\rr{s}\neq\emptyset$. Then
\begin{enumerate}
\item\label{i:tunn-necessity01}
$\Phi(x,X_\rr{s})-H(x)=\Gamma_\rr{m}$
for all 
$x\in X_\rr{m}$;
\item\label{i:tunn-necessity02}
$\Phi(x,X_\rr{s})-H(x)<\Gamma_\rr{m}$
for all $x\in X\setminus(X_\rr{m}\cup X_\rr{s})$.
\end{enumerate}
\end{theorem}

This theorem ensures that $\Gamma_\rr{m}$ is the maximal 
height that must be overcome, starting from any state in the system,
to reach the set $X_\rr{s}$ of ground states of the system.
For this reason the quantity $\Gamma_\rr{m}$ will be also called 
the \definisco{relaxation height} of the system. 

In many situations 
it is of great help 
combining the relaxation height and the stability level 
point of view to get sufficient conditions whose verification 
in the context of specific model is of reasonable difficulty. 
We then state the following theorem.
 
\begin{theorem}
\label{t:schemamnos}
Consider a reversible energy landscape $(X,Q,H,\Delta)$.
Assume $X\setminus X_\rr{s}\neq\emptyset$.
If the not empty set $A\subset X\setminus X_\rr{s}$ and the 
positive real number $a\in\bb{R}_+$ are such that 
\begin{enumerate}
\item\label{i:schemamnosuno}
$\Phi(x,X_\rr{s})-H(x)=a$
for all $x\in A$;
\item\label{i:schemamnosdue}
either
$X\setminus(A\cup X_\rr{s})=\emptyset$
or
$V_x<a$
for all $x\in X\setminus(A\cup X_\rr{s})$;
\end{enumerate}
then
\begin{displaymath}
\Gamma_\rr{m}=a
\;\;\;\textrm{ and }\;\;\;
X_\rr{m}=A
\end{displaymath}
\end{theorem}

We remark that Theorem~\ref{t:tunn} is a corollary 
of Theorem~\ref{t:schemamnos}. Indeed, if the set $A$ and the real number $a$ 
satisfy the hypotheses of Theorem~\ref{t:tunn} then they also satisfy 
the hypotheses of Theorem~\ref{t:schemamnos}. 
Nevertheless, 
since
in the proof of Theorem~\ref{t:schemamnos} the hypothesis that the state space 
$X$ is finite play a crucial role while in the proof of Theorem~\ref{t:tunn}
it does not,
in Section~\ref{s:dimo} the two statements will be proven independently.
The idea in the proof of Theorem~\ref{t:tunn}
could result useful in the study of metastability of Statistical 
Mechanics systems in infinite volume.

\section{Metastable states of Statistical Mechanics lattice models}
\label{s:metlat}
\par\noindent
The theory developed above can be fruitfully applied to study 
metastability in Statistical Mechanics lattice models 
when multiple metastable states are present. 
A natural setup in which this problem can be approached 
is that of Markov chains or Markov processes.
In this context powerful
theories \cite{MNOS,BEGK} have been developed with the aim of 
finding answers valid with maximal
generality and to reduce to a minimum the number of model dependent 
inputs necessary to
describe the metastable behavior of the system.
In this section we shall briefly review the 
{\em pathwise} and the {\em potential theoretic} point of views and 
explain, in this context, the interest of the results stated in 
Section~\ref{s:genris} in order to deal with the case of multiple 
metastable states. 
In particular in Section~\ref{s:potential}
we shall prove a recipe to construct metastable sets in the 
potential theoretic approach sense. 

Consider a finite state space $X$
and a function $q:X\times X\to[0,1]$ called \definisco{connectivity matrix} 
such that for all $x,y\in X$ 
\begin{equation}
\label{conn-prop}
\sum_{w\in X\setminus\{x\}}q(x,w)=1
\;\;\;\;\textrm{ and }\;\;\;\;
q(x,y)=q(y,x)
\end{equation}
Moreover assume that 
for any $x,y\in X$ there exist an integer $n\ge2$ and $x_1,\dots,x_n\in X$ 
such that $x_1=x$, $x_n=y$, and $q(x_i,x_{i+1})>0$ for any 
$i=1,\dots,n-1$. 

Define $Q\subset X\times X$ by letting 
$(x,y)\in Q$ if and only if $q(x,y)>0$.
Consider the \definisco{Hamiltonian} $H:X\to\bb{R}$ and 
the \definisco{cost function}
$\Delta:Q\to\bb{R}_+$ 
such that for all $(x,y)\in Q$ 
\begin{equation}
\label{cost-prop}
H(x)+\Delta(x,y)=\Delta(y,x)+H(y)
\end{equation}
Moreover assume that the following is satisfied: 
(i)
a state 
$x\in X$ is such that $q(x,x)=0$ if and only if 
$\Delta(x,y)=0$ for all $y\in X$ such that $y\neq x$ and $(x,y)\in Q$; 
(ii)
for all $x\in X$ such that $q(x,x)>0$ the cost function is such that 
$\Delta(x,x)=0$.

Given $1/\beta>0$, called \definisco{temperature},
consider the aperiodic ergodic Markov chain on the finite state space 
$X$ and transition matrix $p_\beta:X\times X\to[0,1]$
given by
\begin{equation}
\label{pta-p}
p_\beta(x,y)
:=
q(x,y)e^{-\beta\Delta(x,y)}
\;\;\;
\;\;\;\forall x,y\in X \;\textrm{ and } x\neq y
\end{equation}
and
\begin{equation}
\label{pta-p02}
p_\beta(x,x):=1-\sum_{y\in X:\,x\neq y}p_\beta(x,y)
\;\;\;
\;\;\;\forall x\in X
\end{equation}
From the second equality in (\ref{conn-prop}) and from (\ref{cost-prop})
it follows that the chain is reversible 
with respect to the Gibbs measure 
\begin{equation}
\label{gibbs}
\mu_\beta(\sigma):=\frac{1}{Z_\beta}\,e^{-\beta H(x)}
\end{equation}
where $Z_\beta$ is the partition function, that is to say 
$\mu_\beta(x)p_\beta(x,y)=p_\beta(y,x)\mu_\beta(y)$
for any $x,y\in X$.

Noted that the quaternion $(X,Q,H,\Delta)$ is a reversible 
energy landscape, see Section~\ref{s:rel},
we use the theoretical setup introduced in Section~\ref{s:genris} and,
following \cite{MNOS},
we call 
the set $X_\rr{m}$ introduced in Definition~\ref{t:maxstab}
\definisco{set of metastable states}.
In the following two subsections we shall recall why such a set 
deserves its name.

\subsection{Pathwise approach}
\label{s:path}
\par\noindent
The reason way $X_\rr{m}$ deserves its name has been widely
explained in \cite{MNOS} in the framework of the so called 
{\em pathwise approach} to metastability. 
In that paper the important properties of the states in $X_\rr{m}$ have been 
fully studied in the case of the Metropolis dynamics, namely, when 
the cost function is chosen as 
$\Delta(x,y):=[H(y)-H(x)]_+$, where, for any real $a$, $[a]_+$ is equal
to $a$ if $a\ge0$ and to $0$ otherwise. 
Note that this choice for the function $\Delta$ is coherent with the 
theoretical setup introduced above
since (\ref{cost-prop}) is satisfied.

A partial generalization of the results in \cite{MNOS} to 
the present case, that is to say when 
the function $\Delta$ is as general as explained in Section~\ref{s:genris},
has been done in \cite{CNS01}. The authors had to consider this 
more general situation in \cite{CNS01} since 
such a setup arises 
naturally when Probabilistic Cellular Automata are studied. 

Here, 
in order to justify the name of set of metastable states given 
to $X_\rr{m}$, 
we just recall the following result proven in \cite{CNS01}:
for any $\varepsilon>0$ and $x\in X_\rr{m}$
\begin{equation}
\label{tempo}
\lim_{\beta\to\infty}
\mathbb{P}_x(e^{\beta(\Gamma_\rr{m}-\varepsilon)}
                <\tau_{X_\rr{s}}
              <e^{\beta(\Gamma_\rr{m}+\varepsilon)})
=1
\end{equation}
where $\mathbb{P}_x$ is the probability measure on the space of 
trajectories of the Markov chain started at $x$ and 
$\tau_{X_\rr{s}}$ is the first hitting time 
to $X_\rr{s}$ for the chain started at $x$.
In the case of the Metropolis dynamics, this result has been 
proven in \cite[Theorem~4.1]{MNOS}; in that paper a much more 
detailed description of the metastable behavior of Metropolis dynamics, 
including results in distribution and in law
\cite[Theorems~4.9 and 4.15]{MNOS}, has been given.

One of the remarkable features of the pathwise approach  
is that a {\em constructive} definition of metastable states is given. 
When studying a particular model, in order to find the metastable 
states, one just has to find the set $X_\rr{m}$, 
that is one has to compute the maximum stability level
$\Gamma_\rr{m}$ and identify all
the states whose stability level is equal to $\Gamma_\rr{m}$. 

Unfortunately this is in practice a laborious task. 
Indeed the solution of several variational problems is required. 
It is then very useful the strategy suggested by the 
Theorems~\ref{t:tunn} and \ref{t:schemamnos}
for the computation of 
$\Gamma_\rr{m}$.
This approach has been already used in some situations but always 
in cases in which
there existed a single metastable state,
see for instance \cite[Section~4.2]{MNOS} and \cite[Theorem~2.3]{CNS01}.

\subsection{Potential theoretic approach}
\label{s:potential}
\par\noindent
Different interesting approaches to 
metastability have been developed in the recent literature, see for 
instance \cite{BEGK,BM,BL,BHN,HNT1}. The notion of metastable states 
is given in different ways; but these
different notions are indeed strictly related. For instance in \cite{BEGK}
it is introduced the ``metastable set" which, by using the 
language introduced above, can be reinterpreted in some cases
as $X_\rr{m}\cup X_\rr{s}$. We discuss this issue in this 
subsection in detail.

Following \cite{Bo,BEGK}
we apply the theory developed in the Section~\ref{s:genris}
to the model defined at the beginning of Section~\ref{s:metlat}.
The potential theoretic approach to metastability, introduced in 
\cite{BEGK,Bo}, has been lately developed for
Kawasaki dynamics in \cite{BHN,HNT1} and for probabilistic cellular 
automata in \cite{NS}. 

In this approach the main definition of interest 
is that of metastable set \cite{BEGK}.
For a generic Markov chain 
the \textit{Dirichlet form} is defined as the functional 
\begin{equation}
\label{diri}
\mathfrak{E}_\beta[h]
=\frac{1}{2}\sum_{x,y\in X} 
\mu_\beta(x)p_\beta(x,y) [h(x)-h(y)]^2
=
\frac{1}{2}
\sum_{x,y\in X} 
\frac{1}{Z_\beta}
e^{-\beta[H(x)+\Delta(x,y)]} {[h(x)-h(y)]}^2
\end{equation}
where $h:X\to\bb{R}$ is a generic function and we have used 
(\ref{pta-p}) and the definition (\ref{gibbs}) of Gibbs measure. 

Given two non--empty disjoint sets $A,B\subset X$ the \definisco{capacity} 
of the pair $A$ and $B$ is defined as
\begin{equation}
\label{capac}
\rr{CAP}_\beta(A,B)
:= 
\min_{\newatop{h:X\to[0,1]}{h\vert_A=1,h\vert_B=0}} 
\mathfrak{E}_\beta(h)  
\end{equation}
Note that 
the capacity is a {\em symmetric} function of the sets $A$ and $B$.
It can be proven that the 
right hand side of (\ref{capac}) has a unique minimizer  
called {\em equilibrium potential} of the pair $A$ and $B$ and 
given  by
\begin{displaymath}
h^*_{A,B}(x)=\bb{P}_x(\tau_A<\tau_B)
\end{displaymath}
for any $x\in X$, where $\tau_A$ and $\tau_B$ are, respectively, 
the first hitting time to $A$ and $B$ for the chain started at $x$. 

\begin{definition} 
\label{t:metadef} 
A set $M\subset X$ is said to be 
{\em p.t.a.--metastable}
if 
\begin{equation}
\label{metadef} 
\lim_{\beta\to\infty}
\frac{\max_{x\notin{M}}\mu_\beta(x){[\rr{CAP}_\beta(x,M)]}^{-1}}
     {\min_{x\in{M}}\mu_\beta(x){[\rr{CAP}_\beta(x,M\setminus\{x\})]}^{-1}}
=0
\end{equation}
\end{definition}
The prefix p.t.a. stands for potential theoretic approach. We used 
this expression in order to avoid confusion with the set of 
metastable states $X_\rr{m}$ introduced in Definition~\ref{t:maxstab}.

In \cite{BEGK} and in the related papers the properties of the 
p.t.a.--metastable sets have been widely described. Here we just mention 
the following one to explain why these states are called metastable;
we refer the interested reader to \cite{Bo,BEGK}.
Let $M$ be a p.t.a.--metastable set.
Let $x\in M$ and $J\subset M\setminus\{x\}$ 
be such that for all $y\in M\setminus(J\cup\{x\})$
either 
\begin{displaymath}
\frac{\mu_\beta(y)}{\mu_\beta(x)}\ll1
\;\;\;\;\;\;\textrm{or}\;\;\;\;\;\;
\frac{\rr{CAP}_\beta(y,x)}{\rr{CAP}_\beta(y,J)}\ll1
\end{displaymath}
where by $\ll$ we mean that the ratios on the left--hand sides are 
uniformly bounded from above by a function of $\beta$ tending to 
zero in the limit $\beta\to\infty$.
Then
\begin{displaymath}
\bb{E}_x[\tau_J]=\frac{\mu_\beta(A(x))}{\rr{CAP}(x,J)}(1+o(1))
\end{displaymath}
and
\begin{displaymath}
\bb{P}_{x}(\tau_J>t\bb{E}_x\tau_J)=[1+o(1)]e^{-t[1+o(1)]}\,\,\,t\geq0
\end{displaymath}
for $\beta$ large enough, where $\bb{E}_x$ is the average for the chain started 
at $x$ and, for any $x\in M$, 
the \definisco{valley} $A(x)\subset X$ is defined as 
\begin{displaymath}
A(x):=\{y\in X:\bb{P}_y(\tau_x=\tau_M)=\sup_{z\in M}
\bb{P}_y(\tau_z=\tau_M)\}
\end{displaymath}

When the potential theoretic approach is applied to study metastability in 
specific models the main difficulties are the 
identification of the p.t.a.--metastable sets 
(e.g., the set made by the two configurations with all 
the spins respectively equal to minus one and to plus one in the 
standard Ising model)
and that of giving sharp estimates to the capacities among the 
elements of this set.
In this subsection we will show how the first of the two points 
above can be approached by using the notion of metastable states 
introduced in Definition~\ref{t:maxstab}.

It is important to note that the Definition~\ref{t:maxstab} of set of 
metastable states and the Definition~\ref{t:metadef} of p.t.a.--metastable
sets are different in spirit. 
The set of metastable states $X_\rr{m}$ 
is defined univocally, 
that is to say that all the states that are thought to deserve the name of 
metastable states are collected in the same set $X_\rr{m}$.
In other words Definition~\ref{t:maxstab} is constructive and the 
set of metastable states $X_\rr{m}$ is unique, even though its cardinality 
can be larger than one. 
On the other hand the p.t.a.--metastable 
sets are defined as those sets of states satisfying condition
(\ref{metadef}), so that several 
p.t.a.--metastable sets can exist. 
This subtle, but important, difference is often hidden 
when systems with a unique metastable state ($|X_\rr{m}|=1$) are considered, 
but it emerges in all its importance when multiple metastable 
states ($|X_\rr{m}|\ge2$) are present. 

This aspect of the potential theoretic approach is, indeed, one 
of its distinguishing points. It reveals the high degree of tunability 
of the theory. 
The states in $X_\rr{m}$ are metastable in the sense that 
in order to decrease the energy, starting from them, the 
highest energy barrier (the relaxation height) has to be 
overcome; this is for sure a definition of metastable states
absolutely close to the empirical physical meaning of the word metastable. 
In Theorem~\ref{t:metas} we will show how to construct
p.t.a.--metastable sets by means of states in $X_\rr{m}$ and $X_\rr{s}$;
in some sense we will construct p.t.a--metastable sets by wisely collecting 
``maximally stable" metastable states. 

One could also enlarge its point of view by considering all the 
states such that starting from them the energy barrier that has to be 
overcome in order to lower the energy is larger than or equal to $V$, for 
some $V\in\bb{R}$ such that $0< V<\Gamma_\rr{m}$.
This set would obviously contain $X_\rr{m}$, but, for $V$ small enough, 
also some states not in $X_\rr{m}$ will belong to such a set. 
These states are not ``maximally stable" metastable states, 
but, to some extent, they can be considered metastable since the energy 
barrier $V$ has to be payed in order to decrease the energy. 
The potential theoretical approach is well suited to study also 
these states, indeed one just have to look for the p.t.a.--metastable 
states inside the new set just defined above. 
In the framework of the pathwise approach something 
in this sense has also been proven in \cite{MNOS}, 
see, for instance, the definition \cite[equation~(2.13)]{MNOS} of metastable 
set at a prescribed level and the related result 
\cite[Theorem~3.1]{MNOS}.

In this subsection we try to clarify the connection between the 
``maximally stable" metastable 
sets in the two specified senses and, in particular, we will show 
how to construct the p.t.a.--metastable sets once the set of 
metastable states $X_\rr{m}$ 
is known.
This seems to be a very smart recipe since $X_\rr{m}$ 
can be identified on the basis of the theory developed in 
Section~\ref{s:genris}, for instance by means of the sufficient conditions 
given in Theorem~\ref{t:tunn}
or those in Theorem~\ref{t:schemamnos}.

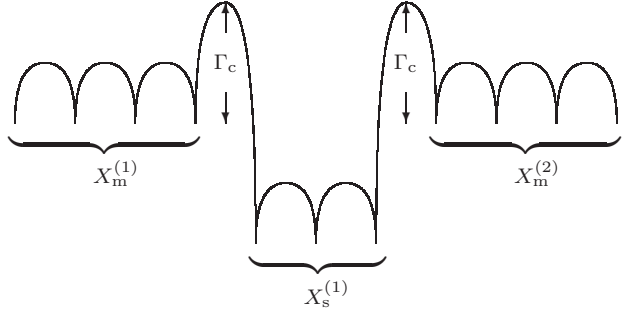
\begin{figure}[t]
 \begin{picture}(100,130)(0,50)
 \thinlines
 \setlength{\unitlength}{0.08cm}
 \qbezier(50,50)(50,60)(55,60)
 \qbezier(55,60)(60,60)(60,50)
 \qbezier(60,50)(60,60)(65,60)
 \qbezier(65,60)(70,60)(70,50)
 \qbezier(70,50)(70,60)(75,60)
 \qbezier(75,60)(80,60)(80,50)
 \qbezier(80,50)(80,70)(85,70)
 \qbezier(85,70)(90,70)(90,30)
 \put(85,65){\vector(0,1){5}}
 \put(85,55){\vector(0,-1){5}}
 \put(83,59){${\scriptstyle \Gamma_\rr{c}}$}
 \qbezier(90,30)(90,40)(95,40)
 \qbezier(95,40)(100,40)(100,30)
 \qbezier(100,30)(100,40)(105,40)
 \qbezier(105,40)(110,40)(110,30)
 \qbezier(110,30)(110,70)(115,70)
 \qbezier(115,70)(120,70)(120,50)
 \qbezier(120,50)(120,60)(125,60)
 \qbezier(125,60)(130,60)(130,50)
 \qbezier(130,50)(130,60)(135,60)
 \qbezier(135,60)(140,60)(140,50)
 \qbezier(140,50)(140,60)(145,60)
 \qbezier(145,60)(150,60)(150,50)
 \put(115,65){\vector(0,1){5}}
 \put(115,55){\vector(0,-1){5}}
 \put(113,59){${\scriptstyle \Gamma_\rr{c}}$}
 \put(89,28){$\underbrace{\phantom{mmmmm}}$}
 \put(98,20){${\scriptstyle X^{(1)}_\rr{s}}$}
 \put(49,48){$\underbrace{\phantom{mmmmmmm}}$}
 \put(63,40){${\scriptstyle X^{(1)}_\rr{m}}$}
 \put(119,48){$\underbrace{\phantom{mmmmmmm}}$}
 \put(133,40){${\scriptstyle X^{(2)}_\rr{m}}$}

 \end{picture}
 \caption{Schematic representation of the energy landscape
          aimed to illustrate the definition of the equivalence 
          relation $\sim$ and the partition in equivalence classes  
          of the sets $X_\rr{m}$ of metastable states and $X_\rr{s}$ of 
          ground states.}
 \label{f:figeq}
\end{figure}

Define on $X$ the following relation:
$x,y\in X$ 
are $x\sim y$ if and only if 
$\Phi(x,y)-H(x)<\Gamma_\rr{m}$
and 
$\Phi(y,x)-H(y)<\Gamma_\rr{m}$.  
It is easy to prove that $\sim$ is an equivalence relation. 
Assume, now, that $X\setminus X_\rr{s}\neq\emptyset$, so that 
$X_\rr{m}$ is not empty. 
We denote by $X^{(1)}_\rr{m},\dots,X^{(k_\rr{m})}_\rr{m}$ 
the equivalence classes in which $X_\rr{m}$ is partitioned with respect to the 
relation $\sim$
(see figure~\ref{f:figeq}).
Analogously, 
we denote by $X^{(1)}_\rr{s},\dots,X^{(k_\rr{s})}_\rr{s}$ 
the equivalence classes in which the relation $\sim$ partitions 
$X_\rr{s}$
(see figure~\ref{f:figeq}).

\begin{theorem}
\label{t:metas}
Assume that $X\setminus X_\rr{s}\neq\emptyset$
and $X\setminus(X_\rr{s}\cup X_\rr{m})\neq\emptyset$.
Chose arbitrarily 
$x_{\rr{s},i}\in X^{(i)}_\rr{s}$ for any $i=1,\dots,k_\rr{s}$
and 
$x_{\rr{m},i}\in X^{(i)}_\rr{m}$ for any $i=1,\dots,k_\rr{m}$.
The set 
\begin{displaymath}
\{x_{\rr{s},1},\dots,x_{\rr{s},k_\rr{s}},
  x_{\rr{m},1},\dots,x_{\rr{m},k_\rr{m}}\}
\end{displaymath}
is p.t.a.--metastable.
\end{theorem}

In words we can say that any set constructed by considering one and 
only one element of each equivalence class $X^{(i)}_\rr{m}$ and 
$X^{(j)}_\rr{s}$ is a p.t.a.--metastable set.

\section{Metastable behavior of the Blume--Capel model}
\label{s:bcmodel}
\par\noindent
In this section we apply the theory described above 
to the specific case of the Blume--Capel model with zero chemical potential.

As it has already been pointed out in the pertinent literature, 
in the study of the metastable behavior of a Statistical Mechanics 
model 
less details on the structure of the energy landscape of the model 
yield less instructive results on the metastable behavior of the system. 
This remark is quite obvious, but nevertheless interesting, since 
the study of the metastable behavior of a system is typically a very 
difficult task. So it is useful to understand fully to which 
extent the description of metastability 
can be pushed forward once some model dependent 
results are known. 

Our results will be then stated according to the following main idea: 
we shall state model dependent lemmas containing informations 
on the energy landscape of the Blume--Capel model followed closely
by theorems stating properties of the metastable states
of the system that can be deduced by using the related lemma. 

\subsection{The Blume--Capel model}
\label{s:bcdef}
\par\noindent
Consider a finite squared torus (periodic boundary 
condition) $\Lambda\subset\bb{Z}^2$ endowed with the 
\definisco{Euclidean} distance $\rr{d}:\Lambda\times\Lambda\to\bb{R}_+$
on the torus. 
As usual we shall misuse the notation and let 
\begin{displaymath}
\rr{d}(I,J):=\min_{i\in I,j\in J}\rr{d}(i,j)
\end{displaymath}
for any $I,J\subset\Lambda$. 
We say that two sites $i,j\in\Lambda$ are \definisco{nearest 
neighbor} if and only if $\rr{d}(i,j)=1$. 

Let $\{-1,0,+1\}$ be the 
single spin state space and $\cc{X}:=\{-1,0,+1\}^\Lambda$ be the 
configuration space. 
The Hamiltonian of the model is 
\begin{equation}
\label{ham-bc}
G(\sigma)
=
\sum_{<i,j>}
(\sigma(i)-\sigma(j))^2
-\lambda\sum_{i\in\Lambda}\sigma(i)^2
-h\sum_{i\in\Lambda}\sigma(i)
\end{equation}
for any $\sigma\in \cc{X}$, 
where the first sum is extended to the pairs of nearest neighbors, 
$\lambda\in\bb{R}$ is called \definisco{chemical potential}, and 
$h\in\bb{R}$ is called \definisco{magnetic field}. 
The metastable states for $h>\lambda>0$ have been widely studied in 
\cite{CO,CNS03}. 

In this paper we shall attack the case $\lambda=0$ 
which is particularly relevant because two metastable 
states will be proven to exist; in other words from now on we shall
consider the zero chemical potential 
Blume--Capel model 
defined by the Hamiltonian
\begin{equation}
\label{ham-bc0}
H(\sigma)
=
\sum_{<i,j>}
(\sigma(i)-\sigma(j))^2
-h\sum_{i\in\Lambda}\sigma(i)
\end{equation}
Given $\sigma\in\cc{X}$, $H(\sigma)$ will be also called \definisco{energy} 
of the configuration $\sigma$. 

The stochastic version of the model is the Markov chain $\sigma_t$, 
with $t=0,1,\dots$ the discrete time variable, with 
transition probabilities (\ref{pta-p}) 
with $\Delta(\sigma,\eta):=[H(\eta)-H(\sigma)]_+$ and 
the connectivity matrix defined as 
\begin{displaymath}
q(\sigma,\eta):=
\left\{
\begin{array}{cl}
0 & \textrm{ if } \sigma,\eta \textrm{ differ at more than one site}
\\
{\displaystyle 
 \frac{1}{2|\Lambda|}
}
 & \textrm{ if } \sigma,\eta \textrm{ differ at one single site}
\end{array}
\right.
\end{displaymath}
for any $\sigma,\eta\in \cc{X}$ and $\sigma\neq\eta$, and $q(\sigma,\sigma)=1$
for any $\sigma\in \cc{X}$. 
Note that, since the dynamics is Metropolis, all the results in \cite{MNOS}
hold for this model.

It is worth nothing that from the definition of the transition matrix
$q$ given above, it follows that for any $\sigma,\eta\in\cc{X}$ 
there exist a path connecting $\sigma$ to $\eta$, 
namely, $|\Omega(\sigma,\eta)|\ge1$.

\subsection{Metastable states of the Blume--Capel model}
\label{s:bcmet}
\par\noindent
We shall discuss the metastable behavior of the zero chemical potential 
Blume--Capel model for the following choice of the parameters.
For any positive real $a$ we let $\lfloor a\rfloor$ be the largest 
integer smaller than or equal to $a$.

\begin{condition}
\label{t:parametri}
The magnetic field $h$ and the torus $\Lambda$ are such that 
$0<h<1$, $\lfloor 2/h\rfloor$ is not integer, and $|\Lambda|\ge49/h^4$ 
finite. 
\end{condition}

For $h>0$
it is immediate 
to remark that the set of ground states of the energy is 
$\cc{X}_\rr{s}=\{\puno\}$,
where $\puno\in \cc{X}$ is the configuration such that 
$\puno(i)=+1$ for all $i\in\Lambda$. 
Indeed the exchange interaction, 
i.e., the positive defined first term of the Hamiltonian in (\ref{ham-bc0}), 
gives its minimal contribution which is equal to zero. 
And so does the magnetic field part. 

Other two very relevant configurations are $\muno$ and 
$\zero$, that is the configuration in which all the spin are 
minus one and the one in which all the spins are zero. 
Indeed in these configurations the exchange part of the energy
is minimal, although the magnetic part is not. We also note that 
\begin{displaymath}
H(\puno)=-|\Lambda|h
<
H(\zero)=0
<
H(\muno)=+|\Lambda|h
\end{displaymath}

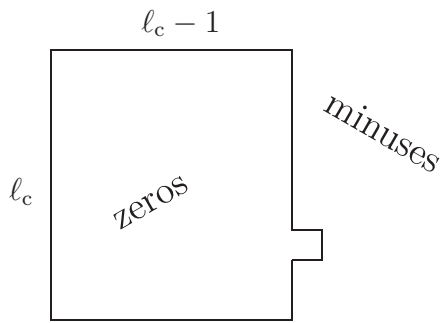
\begin{figure}[t]
 \begin{picture}(100,150)(-50,25)
 \thinlines
 \setlength{\unitlength}{0.08cm}
 \put(60,15){\line(1,0){40}}
 \put(60,15){\line(0,1){45}}
 \put(60,60){\line(1,0){40}}
 \put(100,15){\line(0,1){10}}
 \put(100,60){\line(0,-1){30}}
 \put(100,25){\line(1,0){5}}
 \put(100,30){\line(1,0){5}}
 \put(105,25){\line(0,1){5}}
 \put(75,63){${\ell_\rr{c}-1}$}
 \put(53,35){${\ell_\rr{c}}$}
 \put(70,30){\begin{turn}{30}{\large zeros}\end{turn}}
 \put(105,50){\begin{turn}{-30}{\large minuses}\end{turn}}
 \end{picture}
 \caption{Schematic representation of the configuration $\pcri$. 
          The configuration $\qcri$ has the same geometry with 
          the zeros replaced by the pluses
          and 
          the minuses replaced by the zeros.
          Note that the protuberance can be either on the left 
          or on the right vertical edge (the longest ones in the 
          picture) and there it 
          can be shifted freely.}
 \label{f:fig02}
\end{figure}

On physical grounds, see (for instance) the zero temperature 
phase diagram in \cite[Fig.~1]{CO}, it is quite reasonable to guess that 
$\muno$ and $\zero$ are possible metastable states 
of the system. This is indeed true as it will be stated 
in the following Theorem~\ref{t:meta-bc01}.
To get this result we shall use the strategy outlined 
in Section~\ref{s:genris}: in particular we will 
use the Theorem~\ref{t:schemamnos} with guess set $A=\{\muno,\zero\}$;
this will yield the sufficient 
model dependent statements.

We now define the \definisco{critical length} of the model as 
\begin{equation}
\label{lc}
\ell_\rr{c}:=\big\lfloor\frac{2}{h}\big\rfloor+1
\end{equation}
and remark that, by Condition~\ref{t:parametri}, it follows that 
\begin{equation}
\label{lcp}
\frac{2}{h}<\ell_\rr{c}<\frac{2}{h}+1
\;\;\;\textrm{ and }\;\;\;
\ell_\rr{c}\ge3
\end{equation}
We denote by 
$\pcri$ the set of configurations 
in which all the spins are minus excepted those, which are zeros, in a 
rectangle of sides long $\ell_\rr{c}$ and $\ell_\rr{c}-1$ and 
in a site adjacent to one of the longest sides of the rectangle 
(see figure~\ref{f:fig02}).
We denote by 
$\qcri$ the set of configurations 
in which all the spins are zeros excepted those, which are pluses, in a 
rectangle of sides long $\ell_\rr{c}$ and $\ell_\rr{c}-1$ and 
in a site adjacent to one of the longest sides of the rectangle
(see the caption of figure~\ref{f:fig02}).
From (\ref{ham-bc0}) it follows that
\begin{equation}
\label{gammac0}
H(\pcri)-H(\muno)
\!
=
\!
H(\qcri)-H(\zero)
\!
=
\!
4\ell_\rr{c}-h[\ell_\rr{c}(\ell_\rr{c}-1)+1]
\end{equation}
We then set
\begin{equation}
\label{gammac}
\Gamma_\rr{c}
:=
H(\pcri)-H(\muno)
=
H(\qcri)-H(\zero)
\end{equation}
A simple direct computation shows that for $h$ small one has 
$\Gamma_\rr{c}\sim4/h$.

We are now ready to state the model dependent lemmas on which it will 
be based the description of the metastable behavior 
of the zero chemical potential
Blume--Capel model (\ref{ham-bc0})
(see Section~\ref{s:dimo-meta} for the proof).

\begin{lemma}
\label{t:mdep-bc0.5}
Consider the zero chemical potential Blume--Capel model (\ref{ham-bc0})
and assume that Condition~\ref{t:parametri} is satisfied.
We have that 
\begin{enumerate}
\item\label{i:mdep-bc0.5-01.5}
for any configuration $\eta\in\pcri$ there exists a 
path $\omega\in\Omega(\muno,\puno)$ such that 
$\Phi_\omega-H(\muno)=\Gamma_\rr{c}$, 
$\omega_i=\eta$ for some $i\in\{1,\dots,n\}$ (where $n$ is the length 
of the path), and 
$H(\omega_j)<H(\muno)+\Gamma_\rr{c}$ for all $j\in\{1,\dots,n\}$ and 
$j\neq i$;
\item\label{i:mdep-bc0.5-02.5}
for any configuration $\eta\in\qcri$ there exists a 
path $\omega\in\Omega(\zero,\puno)$ such that 
$\Phi_\omega-H(\zero)=\Gamma_\rr{c}$, 
$\omega_i=\eta$ for some $i\in\{1,\dots,n\}$ (where $n$ is the legth
of the path), and 
$H(\omega_j)<H(\zero)+\Gamma_\rr{c}$ for all $j\in\{1,\dots,n\}$ and 
$j\neq i$.
\end{enumerate}
\end{lemma}

\begin{lemma}
\label{t:mdep-bc01}
Consider the zero chemical potential Blume--Capel model (\ref{ham-bc0})
and assume that Condition~\ref{t:parametri} is satisfied.
We have that 
\begin{enumerate}
\item\label{i:mdep-bc01-01}
$\Phi(\muno,\puno)-H(\muno)
=
\Phi(\zero,\puno)-H(\zero)
=
\Gamma_\rr{c}
$;
\item\label{i:mdep-bc01-02}
$V_\sigma<\Gamma_\rr{c}$ for any 
$\sigma\in \cc{X}\setminus\{\muno,\zero,\puno\}$.
\end{enumerate}
\end{lemma}

Note that the above lemma ensures the validity 
of the hypotheses of Theorem~\ref{t:schemamnos} with 
$A=\{\muno,\zero\}$.
Then, by using the general results 
discussed in the above section, 
one can identify 
the set of metastable states and the relaxation height
(see Section~\ref{s:dimo-meta} for the proofs).

\begin{theorem}
\label{t:meta-bc01}
Consider the zero chemical potential Blume--Capel model (\ref{ham-bc0})
and assume that Condition~\ref{t:parametri} is satisfied.
We have that 
\begin{displaymath}
\cc{X}_\rr{m}=\{\muno,\zero\}
\;\;\textrm{ and }\;\;
\Gamma_\rr{m}=\Gamma_\rr{c}
\end{displaymath}
\end{theorem}

Moreover we can prove a theorem giving the asymptotic of the 
exit time for the system started at the metastable states. 

\begin{theorem}
\label{t:meta-bc02}
Consider the zero chemical potential Blume--Capel model (\ref{ham-bc0})
and assume that Condition~\ref{t:parametri} is satisfied.
The following holds:
\begin{enumerate}
\item\label{i:meta-bc02-01}
considered the chain started at $\muno$, for any $\varepsilon>0$
\begin{displaymath}
\lim_{\beta\to\infty}
\mathbb{P}_\muno(e^{\beta(\Gamma_\rr{c}-\varepsilon)}
                 <\tau_\puno
                 <e^{\beta(\Gamma_\rr{c}+\varepsilon)})
=1
\end{displaymath}
and 
\begin{displaymath}
\lim_{\beta\to\infty}\frac{1}{\beta}\mathbb{E}_\muno[\tau_\puno]=\Gamma_\rr{c}
\end{displaymath}
where $\tau_\puno$ denotes the first hitting 
to $\puno$ of the chain started at $\muno$ and 
$\mathbb{P}_\rr{\muno}$ and 
$\mathbb{E}_\rr{\muno}$ denote, respectively, the probability 
and the average on the space of trajectories started at $\muno$;
\item\label{i:meta-bc02-02}
Considered the chain started at $\zero$, for any $\varepsilon>0$
\begin{displaymath}
\lim_{\beta\to\infty}
\mathbb{P}_\zero(e^{\beta(\Gamma_\rr{c}-\varepsilon)}
                 <\tau_\puno
                 <e^{\beta(\Gamma_\rr{c}+\varepsilon)})
=1
\end{displaymath}
and 
\begin{displaymath}
\lim_{\beta\to\infty}\frac{1}{\beta}\mathbb{E}_\zero[\tau_\puno]=\Gamma_\rr{c}
\end{displaymath}
where, here, $\tau_\puno$ denotes the first hitting 
to $\puno$ of the chain started at $\zero$
and 
$\mathbb{P}_\rr{\zero}$ and 
$\mathbb{E}_\rr{\zero}$ denote, respectively, the probability 
and the average on the space of trajectories started at $\zero$.
\end{enumerate}
\end{theorem}

\subsection{Escape mechanism}
\label{s:nuclea}
\par\noindent
One important result in metastability, besides proving the 
asymptotic on the exit time, is that of identifying the 
escape mechanism. The typical result is that, in order to 
perform the transition to the stable state, the system
has to nucleate a critical droplet of the stable phase 
inside the metastable one. To prove this result supplementary model 
dependent properties are needed.

\begin{lemma}
\label{t:mdep-bc02}
Consider the zero chemical potential Blume--Capel model (\ref{ham-bc0})
and assume that Condition~\ref{t:parametri} is satisfied.
We have that 
\begin{enumerate}
\item\label{i:mdep-bc02-01}
any path $\omega\in\Omega(\muno,\puno)$ such that 
$\Phi_\omega-H(\muno)=\Gamma_\rr{c}$
visits $\pcri$, 
that is there exists an integer $i$ such that $\omega_i\in\pcri$;
\item\label{i:mdep-bc02-02}
any path $\omega\in\Omega(\zero,\puno)$ such that 
$\Phi_\omega-H(\zero)=\Gamma_\rr{c}$
visits $\qcri$, 
that is there exists an integer $i$ such that $\omega_i\in\qcri$.
\end{enumerate}
\end{lemma}

\begin{theorem}
\label{t:meta-bc03}
Consider the zero chemical potential Blume--Capel model (\ref{ham-bc0})
and assume that Condition~\ref{t:parametri} is satisfied.
The following holds:
\begin{enumerate}
\item\label{i:meta-bc03-01}
considered the chain started at $\muno$, we have that
\begin{displaymath}
\lim_{\beta\to\infty}
\mathbb{P}_\muno(\tau_\pcri <\tau_\puno)
=1
\end{displaymath}
where 
$\tau_\pcri$ and $\tau_\puno$ denote respectively the first hitting 
time to $\pcri$ and $\puno$ for the chain started at $\muno$;
\item\label{i:meta-bc03-02}
considered the chain started at $\zero$, we have that
\begin{displaymath}
\lim_{\beta\to\infty}
\mathbb{P}_\zero(\tau_\qcri <\tau_\puno)
=1
\end{displaymath}
where, here, 
$\tau_\qcri$ and $\tau_\puno$ denote respectively the first hitting 
time to $\qcri$ and $\puno$ for the chain started at $\zero$. 
\end{enumerate}
\end{theorem}

\subsection{Remark on the proof of the nucleation property}
\label{s:remark}
\par\noindent
When studying the metastable behavior of a system, the detail 
of the results that one gets depends on the amount of model 
dependent properties that one is able to prove. 
In order to prove the time asymptotic in Theorem~\ref{t:meta-bc02}
only Lemma~\ref{t:mdep-bc01} is needed.
As already remarked, to get the nucleation property in 
Theorem~\ref{t:meta-bc03} it is needed the Lemma~\ref{t:mdep-bc02} 
in which it is identified the state that must 
necessarily be visited by a path joining the 
metastable state to the stable one with maximal height 
equal to the maximal stability level plus the energy of the 
starting metastable state. In particular it is 
proven that such a state is in $\pcri$ for the paths starting at $\muno$ 
and in $\qcri$ for those starting at $\zero$. 

Following the nomenclature in \cite{MNOS}, 
to which we refer the interested reader for 
a more detailed discussion, 
we say that, given $\sigma,\eta\in \cc{X}$, a set $\cc{Y}\subset \cc{X}$ is a 
\definisco{gate} for the pair of configurations $\sigma$ and $\eta$
if and only if 
all the configurations in $\cc{Y}$ have energy equal to the 
communication height between $\sigma$ and $\eta$ and 
any path joining $\sigma$ to $\eta$ 
with maximal height equal to the communication height 
between $\sigma $ and $\eta$ has to pass necessarily through $\cc{Y}$ itself.
More precisely,
a set $\cc{Y}\subset \cc{X}$ is a 
\definisco{gate} for the pair of configurations $\sigma$ and $\eta$
if and only if $H(\zeta)=\Phi(\sigma,\eta)$ for any $\zeta\in\cc{Y}$ 
and for any path $\omega\in\Omega(\sigma,\eta)$ such that 
$\Phi_\omega=\Phi(\sigma,\eta)$ there must be $\zeta\in\cc{Y}$ 
and $i$ integer such that $\omega_i=\zeta$. 
Thus,
Lemma~\ref{t:mdep-bc02} implies that 
the set $\pcri\subset \cc{X}$ is 
a gate for the pair of configurations 
$\muno$ and $\puno$,
whereas
the set $\qcri\subset \cc{X}$ is 
a gate for the pair of configurations 
$\zero$ and $\puno$.

Less details on the structure of the energy landscape of the model 
would have yielded less instructive results. 
Indeed, in the proof of Lemma~\ref{t:mdep-bc01}, in order 
to get the statement we shall 
define the set $\vbim$ of all the configurations 
having precisely $\ell_\rr{c}(\ell_\rr{c}-1)+1$ minus spins. 
As a byproduct of the proof of Lemma~\ref{t:mdep-bc01} 
we shall also get (for free) that any path 
connecting the metastable state $\muno$ to the stable state $\puno$ 
with maximal height equal to the maximal stability level
plus the energy of the 
starting metastable state
must necessarily visit the set $F(\vbim)$
of configurations in $\vbim$ with minimal energy. 
So that, without any supplementary effort with respect to the 
proof of Lemma~\ref{t:mdep-bc01}, we have that $F(\vbim)$ 
is also a gate for the pair of configurations $\muno$ and $\puno$. 
Then, by the theory developed in \cite{MNOS}, we get 
\begin{displaymath}
\lim_{\beta\to\infty}
\mathbb{P}_\muno(\tau_{F(\vbim)} <\tau_\puno)
=1
\end{displaymath}
where the chain is started at $\muno$ and
$\tau_{F(\vbim)}$ and $\tau_\puno$ denote respectively the first hitting 
time to $F(\vbim)$ and $\puno$.

This result can be looked at as a nucleation statement, but 
less precise if compared to that in the first part of 
Theorem~\ref{t:meta-bc03}. This is somehow natural since it has been 
obtained with less information on the structure of the 
energy landscape of the system. In the language of 
\cite{MNOS} both $\pcri$ and $F(\vbim)$ are gates for the 
pair of configurations $\muno$ and $\puno$, but $\pcri$ is a
so called \definisco{minimal} gate. 

A similar discussion can be repeated for the pair of 
configuration $\zero$ and $\puno$ 
with $\vbim$ replaced by $\vbizero$, namely, the set of configurations 
in which all the spins are zeros excepted 
for $\ell_\rr{c}(\ell_\rr{c}-1)+1$ which are pluses.

\subsection{Remarks on gates and minimal gates}
\label{s:remark-mg}
\par\noindent
In \cite[page 604]{MNOS} a gate $\cc{Y}\subset \cc{X}$
for the pair of configurations $\sigma$ and $\eta$
is said to be \definisco{minimal} if and only if for any 
proper subset $\cc{Y}'$ of $\cc{Y}$ there exists 
a path joining $\sigma$ to $\eta$ 
with maximal height equal to the communication height 
between $\sigma$ and $\eta$ 
which does not pass through $\cc{Y}'$.
In other words a gate is minimal if and only if all its 
proper subsets are not gates.
In view of this we have immediately that the gate
$\vbim$ for the pair $\muno$ and $\puno$ 
and 
the gate $\vbizero$ for the pair $\zero$ and $\puno$ 
are not minimal. 
On the other hand from Lemma~\ref{t:mdep-bc0.5}
we have that $\pcri$ and $\qcri$ are minimal gates for the specified 
pair of configurations.

In general minimal gates are not unique. In \cite[Theorem~5.1]{MNOS}
it is proven that the union $\cc{G}(\sigma,\eta)$ of all the minimal gates 
for the pair of configurations $\sigma$ and $\eta$
is 
the set of what the authors call \definisco{essential saddles}
(the interested reader is referred to \cite[page~604]{MNOS} for the 
definition). 
In our model we can prove that $\pcri$ is the unique minimal gate
for the pair $\muno$ and $\puno$
and $\qcri$ is the unique minimal gate
for the pair $\zero$ and $\puno$, so that we can state the following 
theorem.

\begin{theorem}
\label{t:mdep-bc04}
Consider the zero chemical potential Blume--Capel model (\ref{ham-bc0})
and assume that Condition~\ref{t:parametri} is satisfied.
Then
\begin{displaymath}
\cc{G}(\muno,\puno)=\pcri
\;\;\;\textrm{ and }\;\;\;
\cc{G}(\zero,\puno)=\qcri
\end{displaymath}
\end{theorem}

We finally remark that by \cite[Theorem~5.1]{MNOS}
it follows that 
in our model configurations in $\pcri$ (resp.\ $\qcri$) are the 
sole essential saddles for the pair of configurations 
$\muno$ and $\puno$ (resp.\ $\zero$ and $\puno$).

\section{Proof of the general results}
\label{s:dimo}
\par\noindent
In this section we prove the theorems and lemmas concerning 
the general results stated in 
Section~\ref{s:genris} and Section~\ref{s:metlat}.

Proof of Theorem~\ref{t:tunn}.
First of all we note that, if 
$X\setminus(A\cup X_\rr{s})\not=\emptyset$
one has that 
\begin{equation}
\label{dimomia01}
V_x<a
\;\;\;\;\;\;\forall x\in X\setminus(A\cup X_\rr{s})
\end{equation}
Indeed, 
by item~\ref{i:tunn02} in the hypotheses it follows that
there exists a path $\omega\in\Omega(x,X_\rr{s})$ 
such that $\Phi_\omega-H(x)<a$.
Since $H(x)>H(X_\rr{s})$, we have that the same path 
belongs also to $\Omega(x,I_x)$ so that 
$V_x<\Phi_\omega-H(x)$, which proves (\ref{dimomia01}).

We want to prove, now, that 
\begin{equation}
\label{dimomia10}
V_x=a 
\;\;\;\;\;\;\forall x\in A
\end{equation}
The same 
argument developed above and item~\ref{i:tunn01} in the hypothesis
prove that 
\begin{equation}
\label{dimomia0}
V_x\le a
\;\;\;\;\;\;\forall x\in A
\end{equation}
We are now left with the proof of the lower bound
\begin{equation}
\label{dimomia09}
V_x\ge a
\;\;\;\;\;\;\forall x\in A
\end{equation}
Pick $x\in A$ and, by absurdity, 
assume that $V_x<a$.
Then, there exists $x_1\in X$ and $x_1\neq x$
such that $H(x_1)<H(x)$ and a path 
$\omega_1\in\Omega(x,x_1)$ such that
\begin{equation}
\label{dimomia02}
\Phi_{\omega_1}-H(x)<a
\end{equation}
If it were $x_1\in X_\rr{s}$ we would 
immediately get a contradiction, 
since we would have $\omega_1\in\Omega(x,X_\rr{s})$ and hence 
$\Phi(x,X_\rr{s})-H(x)<a$ in contrast 
with the item~\ref{i:tunn01} of the hypothesis.
Assume, finally, that $x_1\in X\setminus X_\rr{s}$. By
items~\ref{i:tunn01} or \ref{i:tunn02} in the hypothesis
we have that there exists a path 
$\omega_2\in\Omega(x_1,X_\rr{s})$ such that 
\begin{equation}
\label{dimomia03}
\Phi_{\omega_2}-H(x_1)\le a
\end{equation}
Now, note that by gluing the path $\omega_2$ to the last configuration 
of $\omega_1$ 
we get the path $\omega_1\omega_2\in\Omega(x,X_\rr{s})$.
Moreover, 
\begin{displaymath}
\Phi_{\omega_1\omega_2}
=
\max\{\Phi_{\omega_1},\Phi_{\omega_2}\}
\end{displaymath}
and hence
\begin{equation}
\label{dimomia04}
\Phi_{\omega_1\omega_2}-H(x)
=
\max\{\Phi_{\omega_1}-H(x),\Phi_{\omega_2}-H(x)\}
\end{equation}
Note, now, that 
\begin{displaymath}
\Phi_{\omega_2}-H(x)
=
\Phi_{\omega_2}-H(x_1)+H(x_1)-H(x)
\end{displaymath}
Since $H(x_1)<H(x)$, from (\ref{dimomia03}) it follows that 
\begin{equation}
\label{dimomia05}
\Phi_{\omega_2}-H(x)<a
\end{equation}
In conclusion, by (\ref{dimomia02}), (\ref{dimomia04}), and 
(\ref{dimomia05}) we have that 
\begin{displaymath}
\Phi_{\omega_1\omega_2}-H(x)<a
\end{displaymath}
which, recalling that $\omega_1\omega_2\in\Omega(x,X_\rr{s})$ 
contradicts item~\ref{i:tunn01} in the hypothesis.
By contradiction we have that $V_x\ge a$ and hence prove 
(\ref{dimomia09}).

By (\ref{dimomia0}) and (\ref{dimomia09}) it eventually follows 
(\ref{dimomia10}).
Recalling (\ref{gamma}) in Definition~\ref{t:maxstab}, 
from (\ref{dimomia10}) and 
(\ref{dimomia01}) 
we get the desired equality 
$\Gamma_\rr{m}=a$.

We finally remark that, since $\Gamma_\rr{m}=a$, 
from Definition~\ref{t:maxstab}
it follows immediately that 
$X_\rr{m}=A$.
Indeed, since $\Gamma_\rr{m}=a$ and $X_\rr{m}$ is the collection of states 
in $X\setminus X_\rr{s}$ such that $V_x=\Gamma_\rr{m}$, from 
items~\ref{i:tunn01} and \ref{i:tunn02} in the hypothesis 
it follows that $X_\rr{m}=A$. 
\qed

\medskip
Proof of Theorem~\ref{t:tunn-necessity}.
By Definition~\ref{t:maxstab} it follows immediately 
that 
\begin{equation}
\label{nec00}
\Phi(x,I_x)-H(x)=\Gamma_\rr{m}
\;\;\;\;\;\forall x\in X_\rr{m}
\end{equation}
and 
\begin{equation}
\label{nec01}
\Phi(x,I_x)-H(x)<\Gamma_\rr{m}
\;\;\;\;\;\forall x\in X\setminus(X_\rr{m}\cup X_\rr{s})
\end{equation}

Proof of item~\ref{i:tunn-necessity01}:
first of all we note that from (\ref{nec00}) it follows immediately 
that 
\begin{equation}
\label{nec03}
\Phi(x,X_\rr{s})-H(x)\ge\Gamma_\rr{m}
\;\;\;\;\;\forall x\in X_\rr{m}
\end{equation}
Indeed, recalling (\ref{communication-set}), we have that 
$X_\rr{s}\subset I_x$ implies 
$\Phi(x,I_x)-H(x)\le \Phi(x,X_\rr{s})-H(x)$.

To prove the upper bound, we pick $x\in X_\rr{m}$ and note that,
by (\ref{nec00}), (\ref{nec01}), and the finiteness of the 
state space $X$, 
we have that 
there exist a sequence of
$n-1$ states $x_1,\dots,x_{n-1}\in X\setminus X_\rr{s}$, 
a ground state $x_n\in X_\rr{s}$, and $n$ paths 
$\omega_i\in\Omega(x_{i-1},x_i)$ with $i=1,\dots,n$ such that 
\begin{displaymath}
H(x_0)>H(x_1)>H(x_2)>\cdots H(x_{n-1})>H(x_n)
\end{displaymath}
and 
\begin{displaymath}
\Phi_{\omega_i}-H(x_{i-1})\le\Gamma_\rr{m}
\end{displaymath}
for all $i=1,\dots,n$, where we have let $x_0=x$.
Note that the bound is not strict because some of the 
states $x_0,x_1,\dots,x_{n-1}$ could belong to $X_\rr{m}$
(at least $x_0=x$ does belong to $X_\rr{m}$).
We then remark that the path $\omega_1\omega_2\cdots\omega_n$ 
obtained by gluing the above paths belongs to 
$\Omega(x,X_\rr{s})$ and 
\begin{equation}
\label{nec04}
\Phi_{\omega_1\omega_2\cdots\omega_n}-H(x)
=
\max_{i=1,\dots,n}[\Phi_{\omega_i}-H(x)]
\end{equation}
For any $i=2,\dots,n$ 
we have that
\begin{equation}
\label{nec06}
\Phi_{\omega_1}-H(x)\le\Gamma_\rr{m}
\;\;\;\textrm{ and }\;\;\;
\Phi_{\omega_i}-H(x)
=
\Phi_{\omega_i}-H(x_{i-1})+H(x_{i-1})-H(x)
<\Gamma_\rr{m}
\end{equation}
since $H(x_{i-1})<X(x)$ for all $i=1,\dots,n$.
In conclusion we have that 
\begin{displaymath}
\Phi_{\omega_1\omega_2\cdots\omega_n}-H(x)\le\Gamma_\rr{m}
\end{displaymath}
which, recalling that $\omega_1\omega_2\cdots\omega_n\in\Omega(x,X_\rr{s})$ 
proves the upper bound 
\begin{equation}
\label{nec05}
\Phi(x,X_\rr{s})-H(x)\le\Gamma_\rr{m}
\;\;\;\;\;\forall x\in X_\rr{m}
\end{equation}
Equations (\ref{nec03}) and (\ref{nec05}) finally yields 
item~\ref{i:tunn-necessity01} of the theorem.

Proof of item~\ref{i:tunn-necessity02}:
we follow the same strategy as that used in the proof of the 
upper bound (\ref{nec05}) above. We just notice that, since 
$x\in X\setminus(X_\rr{m}\cup X_\rr{s})$, in equation (\ref{nec06}) 
the first upper bound is strict, as it follows from (\ref{nec01}). 
\qed

\medskip
Proof of Theorem~\ref{t:schemamnos}.
If 
$X\setminus(A\cup X_\rr{s})\not=\emptyset$,
by following the same strategy described in the proof of 
Theorem~\ref{t:tunn} one easily gets
\begin{equation}
\label{dimmnos01}
V_x\le a
\;\;\;\;\;\;\forall x\in A
\end{equation}

Since one has to prove that $V_x=a$ for any $x\in A$, a 
lower bound is needed. Then, given a generic $x\in A$, 
we assume by absurdity that $V_x<a$. 
Thus, there exists $x_1\in X$ such that $H(x_1)<H(x)$ and a path 
$\omega_1\in\Omega(x,x_1)$ such that
\begin{equation}
\label{dimmnos02}
\Phi_{\omega_1}-H(x)<a
\end{equation}

If it were $x_1\in X_\rr{s}$ we would 
immediately get a contradiction, 
since we would have $\omega_1\in\Omega(x,X_\rr{s})$ and hence 
$\Phi(x,X_\rr{s})-H(x)<a$ in contrast 
with item~\ref{i:schemamnosuno} in the hypotheses of this theorem.

If, on the other hand, 
$x_1\in X\setminus X_\rr{s}$, 
by exploiting the fact that $X_\rr{s}$ is a finite set,
we can 
proceed as above and find a sequence of
$n-2$ states $x_2,\dots,x_{n-1}\in X\setminus X_\rr{s}$, 
a ground state $x_n\in X_\rr{s}$, and $n-1$ paths 
$\omega_i\in\Omega(x_{i-1},x_i)$ with $i=2,\dots,n$ such that 
\begin{displaymath}
H(x_1)>H(x_2)>\cdots H(x_{n-1})>H(x_n)
\end{displaymath}
and 
\begin{displaymath}
\Phi_{\omega_i}-H(x_{i-1})\le a
\end{displaymath}
for all $i=2,\dots,n$.
Note that the bound is not strict because some of the 
configurations $x_1,\dots,x_{n-1}$ could belong to $A$.
We then remark that the path $\omega_1\omega_2\cdots\omega_n$ 
obtained by gluing the above paths belongs to 
$\Omega(x,X_\rr{s})$ and 
\begin{equation}
\label{dimmnos04}
\Phi_{\omega_1\omega_2\cdots\omega_n}-H(x)
=
\max_{i=1,\dots,n}[\Phi_{\omega_i}-H(x)]
\end{equation}
Let $x_0=x$, for any $i=1,\dots,n$ 
we have that
\begin{displaymath}
\Phi_{\omega_i}-H(x)
=
\Phi_{\omega_i}-H(x_{i-1})+H(x_{i-1})-H(x)
<a
\end{displaymath}
since $H(x_{i-1})<X(x)$ for all $i=1,\dots,n$.
In conclusion we have that 
\begin{displaymath}
\Phi_{\omega_1\omega_2\cdots\omega_n}-H(x)<a
\end{displaymath}
which, recalling that $\omega_1\omega_2\cdots\omega_n\in\Omega(x,X_\rr{s})$ 
contradicts item~\ref{i:schemamnosuno} in the hypotheses of this theorem.

We have finally proven that $V_x=a$ for any 
$x\in A$. This, together with item~\ref{i:schemamnosdue}
in the hypotheses of the theorem, 
implies that $\Gamma_\rr{m}=a$.

We finally remark that, since $\Gamma_\rr{m}=a$, 
from Definition~\ref{t:maxstab}
it follows immediately that 
$X_\rr{m}=A$.
Indeed, since $\Gamma_\rr{m}=a$ and $X_\rr{m}$ is the collection of states 
in $X\setminus X_\rr{s}$ such that $V_x=\Gamma_\rr{m}$, from 
items~\ref{i:tunn01} and \ref{i:tunn02} in the hypothesis 
it follows that $X_\rr{m}=A$. 
\qed

We consider now the model in Subsection~\ref{s:potential} 
and prove the Theorem~\ref{t:metas}.
Before proving the theorem we 
recall the following elementary estimate given in \cite{BHN} in the 
context of Kawasaki dynamics.

\begin{lemma}
\label{t:easy}
For every non--empty disjoint sets 
$A,B\subset X$, there exist constants $0<C_1\leq C_2 < \infty$
(depending on $A$ and $B$) such that for all $\beta>0$
\begin{equation}
\label{easy2}
C_1
\le 
e^{\beta\Phi(A,B)}\,Z_\beta\,\rr{CAP}_\beta(A,B)
\le 
C_2
\end{equation}
\end{lemma}

\medskip
Proof of Lemma~\ref{t:easy}: the lemma can be achieved via the 
same argument developed in the proof of \cite[Lemma~3.1.1]{BHN}.
\qed

\medskip
Proof of Theorem~\ref{t:metas}.
Let $M:=\{x_{\rr{s},1},\dots,x_{\rr{s},k_\rr{s}},
          x_{\rr{m},1},\dots,x_{\rr{m},k_\rr{m}}\}$.
Strategy of the proof: we bound the 
numerator in the left--hand side of (\ref{metadef})
from above and the denominator from below. 
These bounds will be sufficient to prove that the ratio 
tends to zero for $\beta\to\infty$.

We first bound the 
numerator in the left--hand side of (\ref{metadef}) from above.
By using the lower bound in (\ref{easy2}) we have that
\begin{equation}
\label{num00}
\mu_\beta(x){[\rr{CAP}_\beta(x,M)]}^{-1}
\leq 
\mu_\beta(x)
\frac{1}{C_1}
e^{\beta\,\Phi(x,M)}
Z_\beta
=
\frac{1}{C_1}\,e^{\beta[\Phi(x,M)-H(x)]}\\
\end{equation}
for any $x\not\in M$.

Let $x\in X\setminus(X_\rr{m}\cup X_\rr{s})$.
Since $M\cap X^{(i)}_\rr{s}\neq\emptyset$ for any $i=1,\dots,k_\rr{s}$, 
item~\ref{i:tunn-necessity02} of Theorem~\ref{t:tunn-necessity} and 
the definition of the equivalence relation $\sim$ 
given in Section~\ref{s:potential}
imply that 
$\Phi(x,M)-H(x)<\Gamma_\rr{m}$.

Let, now, $x\in (X_\rr{s}\cup X_\rr{m})\setminus M$ (note that this 
set can be possibly empty). 
Since $M\cap X^{(i)}_\rr{s}\neq\emptyset$ for any $i=1,\dots,k_\rr{s}$, 
and 
$M\cap X^{(i)}_\rr{m}\neq\emptyset$ for any $i=1,\dots,k_\rr{m}$, 
from the definition of the relation $\sim$ 
given in Section~\ref{s:potential}, we have that 
$\Phi(x,M)-H(x)<\Gamma_\rr{m}$.

These remarks and 
(\ref{num00}) yield that there exist
two real numbers 
$C_3<\infty$ and $\delta>0$ such that 
\begin{equation}
\label{num}
\mu_\beta(x){[\rr{CAP}_\beta(x,M)]}^{-1}
\leq 
C_3 
e^{\beta(\Gamma_m-\delta)}
\end{equation}
for any $x\not\in M$. 

We, now, bound the 
denominator in the left--hand side of (\ref{metadef}) from below.
By using the upper bound in (\ref{easy2}) we have that
for any $x\in M$
\begin{equation}
\label{den00}
\mu_\beta(x){[\rr{CAP}_\beta(x,M\setminus\{x\})]}^{-1}
\ge 
\mu_\beta(\eta)
\frac{1}{C_2}
e^{\beta\,\Phi(x,M\setminus\{x\})}
Z_\beta
=
\frac{1}{C_2}\,e^{\beta[\Phi(x,M\setminus\{x\})-H(x)]}
\end{equation}

First assume $x\in M\cap X_\rr{m}$.
From the definition of the 
equivalence relation $\sim$ and 
the fact that $|M\cap X^{(i)}_\rr{m}|=1$ for any 
$i=1,\dots,k_\rr{m}$, we have that, provided $|M\cap X_\rr{m}|\ge2$ , 
\begin{displaymath}
\Phi(x,(M\setminus\{x\})\cap X_\rr{m})-H(x)\ge\Gamma_\rr{m}
\end{displaymath}
On the other hand, 
since 
$M\cap X^{(i)}_\rr{s}\neq\emptyset$ for any 
$i=1,\dots,k_\rr{m}$, 
from item~\ref{i:tunn-necessity01} in 
Theorem~\ref{t:tunn-necessity} 
and 
from the definition of the 
equivalence relation $\sim$ 
we have that 
\begin{displaymath}
\Phi(x,(M\setminus\{x\})\cap X_\rr{s})-H(x)\ge\Gamma_\rr{m}
\end{displaymath}
Recalling (\ref{communication-set}), 
from the two inequalities above, we have that 
\begin{displaymath}
\Phi(x,M\setminus\{x\})-H(x)
=
\min_{y\in M\setminus\{x\}}\Phi(x,y)-H(x)
\ge\Gamma_\rr{m}
\end{displaymath}
for any $x\in M\cap X_\rr{m}$.

Suppose, now, 
$x\in M\cap X_\rr{s}$. By proceeding as above, 
recalling (\ref{rev02}), and using that $H(X_\rr{s})<H(y)$ 
for any $y\in X_\rr{m}$, we prove that 
\begin{displaymath}
\Phi(x,M\setminus\{x\})-H(x)
\ge\Gamma_\rr{m}
\end{displaymath}
for any $x\in M\cap X_\rr{s}$.

We have eventually proven that 
$\Phi(x,M\setminus\{x\})-H(x)\ge\Gamma_\rr{m}$ for any $x\in M$.
This remark and 
(\ref{den00}) yields the lower bound 
\begin{equation}
\label{den01}
\mu_\beta(x){[\rr{CAP}_\beta(x,M\setminus\{x\})]}^{-1}
\ge 
\frac{1}{C_2}\,e^{\beta\Gamma_\rr{m}}
\end{equation}
for any $x\in M$. 
Hence, from (\ref{num}) and (\ref{den01}) we finally get
\begin{displaymath}
\frac{\max_{x\notin M}\mu_\beta(x)[\rr{CAP}_\beta(x,M)]^{-1}}
     {\min_{x\in M}\mu_\beta(x)[\rr{CAP}_\beta(x,M\setminus\{x\})]^{-1}}
\le 
\frac{C_3}{C_2}\,e^{-\beta\delta}
\end{displaymath}
that completes the proof of the theorem.
\qed

\section{Proof of results concerning the Blume--Capel model}
\label{s:dimo-meta}
\par\noindent
In this subsection we prove the 
Lemmas~\ref{t:mdep-bc0.5} and \ref{t:mdep-bc01} 
and the Theorems~\ref{t:meta-bc01} and \ref{t:meta-bc02}.
Before starting the proof of Lemmas~\ref{t:mdep-bc0.5} and \ref{t:mdep-bc01} 
we state few technical results on \definisco{polyominoes} 
based on the paper \cite{AC} to which we refer the interested reader for 
more details. 

\subsection{Some results on polyominoes}
\label{s:poly}
\par\noindent
Polyominoes have been widely studied both by physicists, to model 
crystal growth, and by combinatorialists. In combinatorics the main 
problem has been that of counting the number of polyominoes 
given their area or their perimeter \cite{BF}.

In the framework of metastability polyominoes play 
an important role and the typical relevant problem 
is that of finding the polyominoes with minimal perimeter and given 
area. 
This problem has been widely studied 
by Alonso and Cerf \cite{AC} both in dimension two and three. 
In the following lemmas and in the 
corollary, we shall summarize some of the properties proven in that paper 
that we shall use in the sequel. 

In order to state the lemma we need some preliminary 
definitions.
Consider the lattice 
$\bb{Z}^2$ embedded in $\bb{R}^2$.
Two sites of $\bb{Z}^2$ are said to be \definisco{nearest neighbor}
if and only if their mutual Euclidean distance is equal to one.
A \definisco{unit square} is a square of area one, whose center belongs to
$\bb{Z}^2$
and whose vertices belong to its dual
$\bb{Z}^2+(1/2,1/2)$.
A \definisco{polyomino} is a finite union
of unit squares.
Note that
a polyomino is thus defined up to a translation as a subset of the 
plane $\bb{R}^2$ in which $\bb{Z}^2$ is embedded. 

The \definisco{area of a polyomino} is the number of its squares.
The \definisco{boundary of a polyomino} is the collection of unit edges
of the dual lattice 
which belong only to one of the unit squares of the polyomino itself. 
The \definisco{perimeter of a polyomino} is the cardinality
of its boundary, namely, the number of unit edges of the dual lattice 
which belong only to one of the unit squares of the polyomino itself. 
In other words the perimeter counts the number of interfaces on 
$\bb{Z}^2$ between the sites inside the polyomino and those outside.

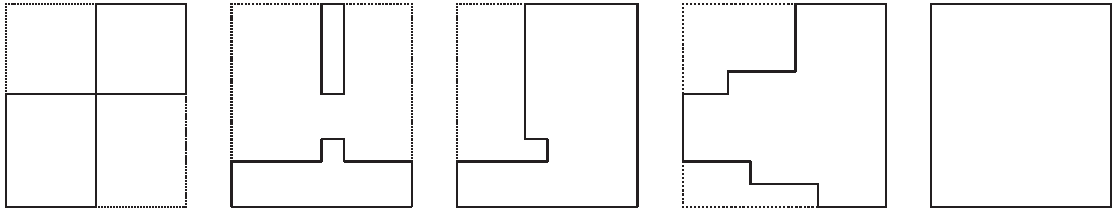
\begin{figure}[t]
 \begin{picture}(100,100)(-95,25)
 \setlength{\unitlength}{0.06cm}
 \thinlines
 \qbezier[30](-40,40)(-40,50)(-40,60)
 \qbezier[30](-40,60)(-30,60)(-20,60)
 \qbezier[40](0,15)(0,27.5)(0,40)
 \qbezier[30](-20,15)(-10,15)(0,15)
 \put(-40,15){\line(1,0){20}}
 \put(-40,40){\line(1,0){20}}
 \put(-40,15){\line(0,1){25}}
 \put(-20,15){\line(0,1){25}}
 \put(-20,40){\line(1,0){20}}
 \put(-20,60){\line(1,0){20}}
 \put(-20,40){\line(0,1){20}}
 \put(0,40){\line(0,1){20}}
 \qbezier[50](10,25)(10,42.5)(10,60)
 \qbezier[50](50,25)(50,42.5)(50,60)
 \qbezier[50](10,60)(30,60)(50,60)
 \put(10,15){\line(1,0){40}}
 \put(10,15){\line(0,1){10}}
 \put(50,15){\line(0,1){10}}
 \put(10,25){\line(1,0){20}}
 \put(50,25){\line(-1,0){15}}
 \put(30,25){\line(0,1){5}}
 \put(35,25){\line(0,1){5}}
 \put(30,30){\line(1,0){5}}
 \put(30,40){\line(1,0){5}}
 \put(30,40){\line(0,1){20}}
 \put(30,60){\line(1,0){5}}
 \put(35,40){\line(0,1){20}}
 \put(60,15){\line(1,0){40}}
 \put(60,15){\line(0,1){10}}
 \put(100,15){\line(0,1){45}}
 \put(60,25){\line(1,0){20}}
 \put(80,25){\line(0,1){5}}
 \put(80,30){\line(-1,0){5}}
 \put(75,30){\line(0,1){30}}
 \put(75,60){\line(1,0){25}}
 \qbezier[50](60,25)(60,42.5)(60,60)
 \qbezier[20](60,60)(67.5,60)(75,60)
 \qbezier[50](110,15)(132.5,15)(155,15)
 \qbezier[50](110,60)(132.5,60)(155,60)
 \qbezier[50](110,15)(110,37.5)(110,60)
 \put(110,25){\line(0,1){15}}
 \put(110,40){\line(1,0){10}}
 \put(120,40){\line(0,1){5}}
 \put(120,45){\line(1,0){15}}
 \put(135,45){\line(0,1){15}}
 \put(135,60){\line(1,0){20}}
 \put(155,60){\line(0,-1){45}}
 \put(155,15){\line(-1,0){15}}
 \put(140,15){\line(0,1){5}}
 \put(140,20){\line(-1,0){15}}
 \put(125,20){\line(0,1){5}}
 \put(125,25){\line(-1,0){15}}
 \put(165,15){\line(1,0){40}}
 \put(165,15){\line(0,1){45}}
 \put(165,60){\line(1,0){40}}
 \put(205,15){\line(0,1){45}}
 \end{picture}
 \caption{From the left to the right: two not connected convex polyominoes,
          a connected not convex polyomino, 
          a connected convex polyomino 
          which is not a convex subset of $\bb{R}^2$, 
          a connected convex polyomino 
          which is a convex subset of $\bb{R}^2$.
          The dotted lines denote the smallest surrounding rectangle.}
 \label{f:poly}
\end{figure}

A polyomino is \definisco{connected} if and only if 
its interior\footnote{The centers of the unit squares forming a 
connected polyomino are a 
nearest neighbor connected subset of $\bb{Z}^2$.
As in \cite{AC} here a polyomino is not necessarily connected, 
different definitions, see for instance 
\cite{BF}, can be found in the literature.}
is a connected subset of $\bb{R}^2$.
A polyomino is \definisco{convex} if and only if its intersection 
with any line parallel to the coordinate axes of $\bb{Z}^2$ is convex.

A polyomino is \definisco{monotone} if and only if its perimeter is
equal to that of its smallest surrounding rectangle.
A polyomino is \definisco{minimal} if and only if any other 
polyomino with the same area has perimeter greater or equal to 
that of the polyomino itself. 

\begin{lemma}
\label{t:polyomini0}
The following holds:
\begin{enumerate}
\item
\label{i:polyomini0-01}
the perimeter of a connected polyomino is greater than or equal 
to that of its smallest surrounding rectangle;
\item
\label{i:polyomini0-02}
a connected polyomino is convex if and only if it is monotone;
\item
\label{i:polyomini0-03}
a minimal polyomino is connected and convex.
\end{enumerate}
\end{lemma}

\medskip
Proof of Lemma~\ref{t:polyomini0}.
Call \definisco{vertical} and \definisco{horizontal} the two 
directions defined by the coordinate axes of $\bb{Z}^2$. 
Define, also, the notion of \definisco{left}, \definisco{right}, 
\definisco{top}, and \definisco{bottom}.
Let a \definisco{column} (resp.\ \definisco{row}) 
be the subset of $\bb{R}^2$ obtained 
by considering the union of all the unit squares whose center belong to 
the same vertical (resp.\ horizontal) line.

Item~\ref{i:polyomini0-01}:
since, by hypothesis, the polyomino is a connected subset of $\bb{R}^2$, 
each row and each column intersecting the smallest rectangle surrounding 
the polyomino
intersects the boundary 
of the polyomino in at least two unit segments of the dual 
lattice. This remark yields the statement. 

Item~\ref{i:polyomini0-02}:
by definition of convex polyominoes
it follows that a connected 
polyomino is convex if and only if 
each vertical and each horizontal 
line passing through the sites of $\bb{Z}^2$ intersects the polyomino 
in a convex not empty subset of the line itself. 
Thus, 
a polyomino is convex if and only if 
each vertical and each horizontal 
line passing through the sites of $\bb{Z}^2$ 
intersects the boundary 
of the polyomino in exactly two unit segments of the dual 
lattice. This remark yields the statement. 

Item~\ref{i:polyomini0-03}:
the fact that a minimal polyomino is connected is quite obvious. 
Indeed, it is sufficient to identify its maximal connected components and 
dispose them in a chain according to the following rule:
excepted for the last component in the chain, 
the right part of the boundary of each 
component
intersects the left part of the boundary of the  
component following it in the chain 
at least on a unit segment of the dual lattice. 
This construction yields a connected polyomino whose perimeter 
is smaller than the perimeter of the original polyomino 
by a quantity which at least equal to twice the number of maximal connected 
components minus one. 

We now prove that a minimal polyomino is convex.
Consider a minimal polyomino $c$ and, by absurdity, assume it is not 
convex. We shall construct another polyomino with the same area
and smaller perimeter. This will yield the statement by 
contradiction. 

Since the polyomino is connected, 
by items~\ref{i:polyomini0-01} and \ref{i:polyomini0-02} 
of this lemma we have that the perimeter $P$ of the 
polyomino $c$ is strictly greater than the perimeter $R$ of its 
smallest surrounding rectangle $r$, that is to say $P>R$.

As a main tool in the sequel of the 
proof we shall use horizontal and vertical 
projections as those introduced in \cite[Section~2, page 6]{AC}, see also 
\cite[Figure~8]{HNT}. 
We first construct a second polyomino $c_1$ by projecting $c$ 
vertically to its bottom, 
that is we consider the polyomino $c_1$ such that 
(i)
the number of unit cell in each column is equal 
to that of the polyomino $c$, 
(ii) the intersection between $c_1$ 
and any column is a convex polyomino, 
and
(iii) the boundary of the polyomino obtained by 
intersecting $c_1$ 
and any column intersects the bottom 
horizontal edge of $r$.
We then construct a third polyomino $c_2$ by projecting $c_1$ 
horizontally to its left, 
that is we consider the polyomino $c_2$ such that 
(i)
the number of unit cell in each row is equal 
to that of the polyomino $c_1$, 
(ii) the intersection between $c_2$ 
and any row is a convex polyomino, 
and
(iii) the boundary of the polyomino obtained by 
intersecting $c_2$ 
and any row intersects the left
vertical edge of $r$.

We note, now, that $c_2$ is connected and convex and that 
the smallest rectangle $r_2$ surrounding $c_2$ 
is a subset of $r$. Since $P>R$, the perimeter $P_2$ of $c_2$ 
is equal to the perimeter of $r_2$, and $r_2\subset r$, we have
we have that $P>R\ge P_2$. 
Which is an absurd since 
the area of $c$ is equal to that of $c_2$.
\qed

\begin{lemma}
\label{t:polyomini01}
For any $n$ positive integer
there exists two positive integers $s$ and $k$, with 
$0\le k<s$, such that 
either (i) $n=s(s-1)+k$ or (ii) $n=s^2+k$.
The set of polyominoes of area $n$ and 
minimal perimeter contains 
a rectangle of side lengths $s$ and $s-1$ with a bar long $k$ 
attached to one of its longest sides in the case (i)
and 
a square of side length $s$ with a bar long $k$
attached to one of its sides
in the case (ii).
\end{lemma}

\medskip
Proof of Lemma~\ref{t:polyomini01}.
The lemma is nothing but a simple restatement of Theorem~2.2 in \cite{AC}.
\qed

\begin{corollary}
\label{t:polyomini02}
For any $n$ positive integer, the square of the perimeter of the polyominoes 
of area $n$ is bounded from below by $16n$.
\end{corollary}

\medskip
Proof of Corollary~\ref{t:polyomini02}.
In the case (i) of Lemma~\ref{t:polyomini01} we have 
that the perimeter of the polyomino is $P=4s$.
Since the polyomino is contained in a square of side length $s$, 
we have that $n\le s^2$. Hence $P^2=16s^2\ge16 n$.

In the case (ii) of Lemma~\ref{t:polyomini01} we have 
that the perimeter of the polyomino is $P=4s+2$. 
Since the polyomino is contained in a rectangle of side lengths $s$ and 
$s+1$, 
we have that $n\le s^2+s$. 
Since
\begin{displaymath}
P^2=(4s+2)^2=16(s^2+s)+4\ge16n+4
\end{displaymath}
the proof of the corollary is completed.
\qed

\subsection{Metastable states of the Blume--Capel model}
\label{s:pmetlat}
\par\noindent
In this subsection we prove the 
Lemmas~\ref{t:mdep-bc0.5} and \ref{t:mdep-bc01} 
and the Theorems~\ref{t:meta-bc01} and \ref{t:meta-bc02}
which allow to identify the metastable states of the Blume--Capel model 
and to deduce the exponential estimate of the exit time. 

\medskip
Proof of Lemma~\ref{t:mdep-bc0.5}.
We construct the following path. 
The path is defined by the following simple rule
\begin{itemize}
\item[--]
start from $\muno$;
\item[--]
transform one minus spin in zero;
\item[--]
iteratively let the zero cluster grow until $\zero$ is 
reached by adding a zero 
protuberance to the longest side of a rectangle and then 
filling the slice with zeros;
\item[--]
transform one zero spin in plus;
\item[--]
iteratively let the plus cluster grow until $\puno$ is 
reached by adding a plus
protuberance to the longest side of a rectangle and then 
filling the slice with pluses.
\end{itemize}
Note that the rectangles that are drawn during the 
above iterative procedure are squares or quasi--squares
(rectangles whose side legths differ by one). 

The path, see figure~\ref{f:fig03}, 
starts at $\muno$ and, passing through $\pcri$ reaches $\zero$,
then passes through $\qcri$ and, finally, reaches $\puno$. 
This path can be divided into two parts 
$\omega_1\in\Omega(\muno,\zero)$ connecting $\muno$ to $\zero$ 
and 
$\omega_2\in\Omega(\zero,\puno)$ connecting $\zero$ to $\puno$. 
By direct inspection, it is very easy to show that 
\begin{displaymath}
\Phi_{\omega_1}-H(\muno)
=
\Phi_{\omega_2}-H(\zero)
=
\Gamma_\mathrm{c}
\end{displaymath}
and that $\omega_1$ and $\omega_2$ attain their maximal height 
respectively only in the states in $\pcri$ and $\qcri$.
This yields the lemma.
\qed

\begin{figure}
 \begin{picture}(300,150)(-90,10)
 \thinlines
 \put(-10,-5){${\scriptstyle \muno}$}
 \put(0,0){\circle*{2}}
 \put(0,0){\line(1,2){20}}
 \put(20,40){\circle*{2}}
 \put(10,40){${\scriptscriptstyle 0}$}
 \put(15,15){${\scriptscriptstyle 4-h}$}
 \put(20,40){\line(1,2){10}}
 \put(30,60){\circle*{2}}
 \put(20,60){${\scriptscriptstyle 0}$}
 \put(15,60){${\scriptscriptstyle 0}$}
 \put(30,45){${\scriptstyle 2-h}$}
 \put(30,60){\line(1,2){10}}
 \put(40,80){\circle*{2}}
 \put(30,80){${\scriptscriptstyle 0}$}
 \put(25,80){${\scriptscriptstyle 0}$}
 \put(25,85){${\scriptscriptstyle 0}$}
 \qbezier(40,80)(41,78)(42.5,75)
 \put(42.5,75){\circle*{2}}
 \put(42.5,70){${\scriptscriptstyle 0}$}
 \put(47.5,70){${\scriptscriptstyle 0}$}
 \put(42.5,65){${\scriptscriptstyle 0}$}
 \put(47.5,65){${\scriptscriptstyle 0}$}
 \put(42.5,75){\line(1,2){10}}
 \put(52.5,95){\circle*{2}}
 \put(42.5,95){${\scriptscriptstyle 0}$}
 \put(37.5,95){${\scriptscriptstyle 0}$}
 \put(42.5,100){${\scriptscriptstyle 0}$}
 \put(37.5,100){${\scriptscriptstyle 0}$}
 \put(37.5,105){${\scriptscriptstyle 0}$}
 \qbezier(52.5,95)(53.5,93)(55,90)
 \put(55,90){\circle*{2}}
 \put(55,85){${\scriptscriptstyle 0}$}
 \put(60,85){${\scriptscriptstyle 0}$}
 \put(55,80){${\scriptscriptstyle 0}$}
 \put(60,80){${\scriptscriptstyle 0}$}
 \put(55,75){${\scriptscriptstyle 0}$}
 \put(60,75){${\scriptscriptstyle 0}$}
 \put(55,90){\line(1,2){10}}
 \put(65,110){\circle*{2}}
 \put(55,110){${\scriptscriptstyle 0}$}
 \put(50,110){${\scriptscriptstyle 0}$}
 \put(55,115){${\scriptscriptstyle 0}$}
 \put(50,115){${\scriptscriptstyle 0}$}
 \put(55,120){${\scriptscriptstyle 0}$}
 \put(50,120){${\scriptscriptstyle 0}$}
 \put(45,120){${\scriptscriptstyle 0}$}
 \qbezier(65,110)(66,109)(67.5,105)
 \put(67.5,105){\circle*{2}}
 \qbezier(67.5,105)(68.5,103)(70,100)
 \put(70,100){\circle*{2}}
 \put(70,95){${\scriptscriptstyle 0}$}
 \put(75,95){${\scriptscriptstyle 0}$}
 \put(80,95){${\scriptscriptstyle 0}$}
 \put(70,90){${\scriptscriptstyle 0}$}
 \put(75,90){${\scriptscriptstyle 0}$}
 \put(80,90){${\scriptscriptstyle 0}$}
 \put(70,85){${\scriptscriptstyle 0}$}
 \put(75,85){${\scriptscriptstyle 0}$}
 \put(80,85){${\scriptscriptstyle 0}$}
 \put(70,100){\line(1,2){5}}
 \qbezier[20](75,110)(80,120)(85,130)
 \qbezier[100](43,80)(150,80)(80,40)
 \put(69,35){${\scriptstyle -h}$}
 \qbezier[20](150,85)(155,75)(160,65)
 \put(160,65){\circle*{2}}
 \put(150,60){\line(1,0){20}}
 \put(150,40){\line(1,0){20}}
 \put(150,40){\line(0,1){20}}
 \put(170,40){\line(0,1){20}}
 \put(152,33){${\scriptstyle \ell_{\textrm{c}}-1}$}
 \put(160,65){\line(1,2){20}}
 \put(180,105){\circle*{2}}
 \put(170,110){\line(1,0){20}}
 \put(170,110){\line(0,1){20}}
 \put(190,110){\line(0,1){20}}
 \put(170,130){\line(1,0){10}}
 \put(185,130){\line(1,0){5}}
 \put(180,130){\line(0,1){5}}
 \put(185,130){\line(0,1){5}}
 \put(180,135){\line(1,0){5}}
 \qbezier(180,105)(181,103)(186,93)
 \put(186,93){\circle*{2}}
 \qbezier[20](186,93)(189.5,86)(193,79)
 \put(190,85){\circle*{2}}
 \qbezier(190,85)(191,84)(196,73)
 \put(196,73){\circle*{2}}
 \put(190,68){\line(1,0){20}}
 \put(190,43){\line(1,0){20}}
 \put(190,43){\line(0,1){25}}
 \put(210,43){\line(0,1){25}}
 \put(213,48){${\scriptstyle \ell_{\textrm{c}}}$}
 \put(192,35){${\scriptstyle \ell_{\textrm{c}}-1}$}
 \put(196,73){\line(1,2){20}}
 \put(216,113){\circle*{2}}
 \put(206,118){\line(1,0){20}}
 \put(206,143){\line(1,0){20}}
 \put(206,118){\line(0,1){25}}
 \put(226,118){\line(0,1){5}}
 \put(226,128){\line(0,1){15}}
 \put(226,128){\line(1,0){5}}
 \put(226,123){\line(1,0){5}}
 \put(231,123){\line(0,1){5}}
 \put(231,133){${\scriptstyle\pcri}$}
 \qbezier(216,113)(217,111)(222,101)
 \put(222,101){\circle*{2}}
 \qbezier[20](222,101)(228,93)(231,83)
 \put(231,83){\circle*{2}}
 \qbezier(231,83)(232,81)(237,71)
 \put(237,71){\circle*{2}}
 \put(232,67){\line(1,0){25}}
 \put(232,42){\line(1,0){25}}
 \put(232,42){\line(0,1){25}}
 \put(257,42){\line(0,1){25}}
 \put(242,34){${\scriptstyle \ell_{\textrm{c}}}$}
 \put(260,47){${\scriptstyle \ell_{\textrm{c}}}$}
 \put(237,71){\line(1,2){20}}
 \put(257,111){\circle*{2}}
 \put(250,116){\line(1,0){25}}
 \put(250,141){\line(1,0){25}}
 \put(250,116){\line(0,1){25}}
 \put(275,116){\line(0,1){5}}
 \put(275,126){\line(0,1){15}}
 \put(275,121){\line(1,0){5}}
 \put(275,126){\line(1,0){5}}
 \put(280,121){\line(0,1){5}}
 \qbezier[20](257,111)(260,105)(263,99)
 \put(152,90){${\scriptstyle 2-h}$}
 \put(185,100){${\scriptstyle -h}$}
 \end{picture}
 \vskip 0.5 cm
 \caption{Schematic representation of path introduced 
          at the beginning of the proof of Lemma~\ref{t:mdep-bc0.5}.
          Only the first part of the path, 
          the one going from $\muno$ to few steps after 
          $\pcri$, is shown.}
 \label{f:fig03}
\end{figure}

Now, we introduce the set 
$\vbim\subset \cc{X}$ as the set of states in $\cc{X}$ such that 
the number of minus spins is equal to 
\begin{displaymath}
|\Lambda|-[\ell_\rr{c}(\ell_\rr{c}-1)+1]
\end{displaymath}
In the following lemma we characterize the set $F(\vbim)$ of the 
minima of the energy of $\vbim$.

\begin{lemma}
\label{t:vbim}
Consider the zero chemical potential Blume--Capel model (\ref{ham-bc0})
and assume that Condition~\ref{t:parametri} is satisfied.
Let $\sigma\in\vbim$ and $N_\sigma\subset\Lambda$ be the 
set of sites $i\in\Lambda$ such that $\sigma(i)\neq-1$.
We have that 
\begin{enumerate}
\item\label{i:vbim01}
the set $N_\sigma$ is not a nearest neighbor connected 
subset of $\Lambda$ winding around the torus $\Lambda$;
\item\label{i:vbim02}
if $\sigma\in F(\vbim)$ then $\sigma(i)=0$ for any $i\in N_\sigma$;
\item\label{i:vbim03}
$\vbim\supset\pcri$;
\item\label{i:vbim04}
$F(\vbim)\supset\pcri$;
\item\label{i:vbim05}
$H(F(\vbim))=H(\muno)+\Gamma_\rr{c}$.
\end{enumerate}
\end{lemma}

\medskip
Proof of Lemma~\ref{t:vbim}.\
Item~\ref{i:vbim01}:
recalling (\ref{lcp}) we have that 
\begin{displaymath}
|N_\sigma|
=
\ell_\rr{c}(\ell_\rr{c}-1)+1
<
\Big(\frac{2}{h}+1\Big)\frac{2}{h}+1
=\frac{4}{h^2}+\frac{2}{h}+1
\le
\frac{4}{h^2}+\frac{2}{h^2}+\frac{1}{h^2}
\end{displaymath}
where in the last inequality 
we have used that, by Condition~\ref{t:parametri}, we have $h<1$. 
The item follows since, by Condition~\ref{t:parametri},
we have that $|\Lambda|\ge49/h^4$. 

Item~\ref{i:vbim02}:
let $\sigma\in\vbim$ and let $r,\ell,m$ not negative integers
such that $r$ is the number of pluses in $\sigma$, 
$\ell$ is the number of plus--minus interfaces, 
and
$m$ is the number of plus--zero interfaces.
Let $\sigma'$ be the configuration obtained 
by flipping to zero the pluses in 
$\sigma$. By (\ref{ham-bc0}), it follows that
\begin{displaymath}
H(\sigma')-H(\sigma)
=
rh-3\ell-m
\le
rh-(\ell+m)
\end{displaymath}

Let $C(\sigma)$ be the polyomino 
obtained by collecting
all the unit squares associated with a plus spin of $\sigma$.
Note that its perimeter is equal to $\ell+m$ and its area to $r$.
Since the polyomino $C(\sigma)$ does not wind around the torus, 
we can apply Corollary~\ref{t:polyomini02} and obtain that 
$(\ell+m)^2\ge16r$.
This bound on the number of interfaces between the pluses and the 
other two types of spins implies that
\begin{displaymath}
H(\sigma')-H(\sigma)
\le
rh-4\sqrt{r}
\end{displaymath}

By studying the parabola of algebraic equation $y=hx^2-4x$ it is 
immediate to show that $H(\sigma')-H(\sigma)\le0$ provided 
$\sqrt{r}<4/h$.
This condition is easily proven to be true by using that 
$r\le\ell_\rr{c}(\ell_\rr{c}-1)+1$. Indeed, by performing the 
same estimate as in the proof of item~\ref{i:vbim01} above, 
we have that 
\begin{displaymath}
r
\le
\ell_\rr{c}(\ell_\rr{c}-1)+1
<
\frac{7}{h^2}
\end{displaymath}
which yields the desired bound
$\sqrt{r}<4/h$.

We can finally conclude that $H(\sigma')\le H(\sigma)$, which completes 
the proof of the item. 
Note that 
the bound is not strict because, in principle, we could have no plus 
spin in $\sigma$, that is to say $r=0$.

Item~\ref{i:vbim03}:
since any configuration in 
$\pcri$ has $|\Lambda|-[\ell_\rr{c}(\ell_\rr{c}-1)+1]$ minus spins,
it follows that 
$\pcri\subset\vbim$.

Item~\ref{i:vbim04}:
for any $\sigma\in F(\vbim)$ item~\ref{i:vbim02} above 
states that the spins not equal to minus are equal to zero. 
Thus, $F(\vbim)$ is a subset $\vbimzero\subset\vbim$ 
made of those configurations 
such that 
all the spins that are not minus are zero. 
Consider $\sigma\in\vbimzero$ and 
denote by $C(\sigma)$ the 
polyomino obtained by 
collecting all the unit squares associated with a spin zero.
Remarked that 
the area of the polyomino $C(\sigma)$ 
is equal to $[\ell_\rr{c}(\ell_\rr{c}-1)+1]$,
by (\ref{ham-bc0}) we have that 
\begin{displaymath}
H(\sigma)-H(\muno)
=\rr{Perimeter}(C(\sigma))-h[\ell_\rr{c}(\ell_\rr{c}-1)+1]
\end{displaymath}
This quantity is minimal if and only if the perimeter of the polyomino 
$C(\sigma)$ is minimal. 
We can then conclude that 
the configurations in $F(\vbim)$ are 
made of 
$[\ell_\rr{c}(\ell_\rr{c}-1)+1]$ zeros in the sea of minuses and 
that the polyomino obtained by 
collecting all the unit squares associated with a zero spin has 
area 
$[\ell_\rr{c}(\ell_\rr{c}-1)+1]$
and minimal perimeter.

Since item~\ref{i:vbim01} above ensures that for any $\sigma\in\vbimzero$
the polyomino $C(\sigma)$ does not wind around the torus, 
we can then apply Lemma~\ref{t:polyomini01} (case (i))
and conclude that $\pcri\subset F(\vbim)$.

Item~\ref{i:vbim05}:
immediate consequence of item~\ref{i:vbim04} above and the definition 
(\ref{gammac}) of $\Gamma_\rr{c}$.
\qed

\medskip
Proof of item~\ref{i:mdep-bc01-01}
of Lemma~\ref{t:mdep-bc01}.
We have to compute 
$\Phi(\muno,\puno)-H(\muno)$
and
$\Phi(\zero,\puno)-H(\zero)$.
In order to show that they are both equal to $\Gamma_\rr{c}$ 
we show first that $\Gamma_\rr{c}$ is an upper bound for 
both. 
This can be done easily, indeed by Lemma~\ref{t:mdep-bc0.5} 
it follows 
\begin{equation}
\label{dasopra}
\Phi(\muno,\puno)-H(\muno)
\le
\Gamma_\mathrm{c}
\;\textrm{ and }\;
\Phi(\zero,\puno)-H(\zero)
\le
\Gamma_\mathrm{c}
\end{equation}

In order to get the lower bound the two problems have to 
be treated separately. 
We focus, first, on the problem $\Phi(\muno,\puno)$ and 
proceed as follows. 

Recall the definition of $\vbim$ given at the beginning of this 
subsection. 
Since the connectivity matrix $q(\sigma,\eta)$ is different from zero
only if the two configurations $\sigma$ and $\eta$ differ for 
the value of one single spin, we have that 
for any not negative integer $n$ smaller than $|\Lambda|$, 
any path in $\Omega(\muno,\puno)$ must necessarily visit
the set of configurations with $n$ minus spins. It then 
follows that 
any path in $\Omega(\muno,\puno)$ must necessarily pass 
through $\vbim$.

Item~\ref{i:vbim05} in Lemma~\ref{t:vbim} states 
that $H(F(\vbim))=H(\muno)+\Gamma_\rr{c}$.
The above remarks imply that any path $\omega\in\Omega(\muno,\puno)$ 
is such that $\Phi_\omega\ge H(F(\vbim))=H(\muno)+\Gamma_\rr{c}$, 
which yields the desired lower bound 
$\Phi(\muno,\puno)-H(\muno)\ge\Gamma_\rr{c}$.

The lower bound 
$\Phi(\zero,\puno)-H(\zero)\ge H(\qcri)-H(\zero)=\Gamma_\rr{c}$
can be achieved, ``mutatis mutandis," with the same argument. 
We do not enter into the 
details, we just remark that the manifold $\vbim$ must be 
replaced by $\vbizero$ defined as 
the set of configurations in which all the spins are zeros excepted 
for $\ell_\rr{c}(\ell_\rr{c}-1)+1$ which are pluses.
\qed

Before starting the proof of 
item~\ref{i:mdep-bc01-02}
of Lemma~\ref{t:mdep-bc01}
we need to state a technical lemma on the zero chemical 
potential Blume--Capel Hamiltonian
ensuring that  
the energy (\ref{ham-bc0}) of a configuration 
is decreased if
a minus spin having at most two minuses between its nearest 
neighbors or a minus spin having three neighboring minus and a 
neighboring plus is flipped to zero.
Moreover, the lemma states that the energy of a configuration 
is decreased if a zero spin with at least two pluses and no minus 
among its nearest neighbors is flipped to plus.

For any $\sigma\in\cc{X}$, $i\in\Lambda$, and $a\in\{-1,0,+1\}$ 
we let $\sigma^{i,a}$ be the configuration such that 
$\sigma^{i,a}(j)=\sigma(j)$ for all $j\in\Lambda$ and $j\neq i$ and 
$\sigma^{i,a}(i)=a$. Note that $\sigma^{i,a}$ differs from $\sigma$ 
by at most the value of the spin associated with the site $i$. 

\begin{lemma}
\label{t:bcflip}
Consider the zero chemical potential Blume--Capel model (\ref{ham-bc0})
and assume that Condition~\ref{t:parametri} is satisfied.
Let $\sigma\in\cc{X}$
\begin{enumerate}
\item\label{i:bcflip01}
if there exists a site $i$ such that $\sigma(i)=-1$, three 
of the nearest neighbors of $i$ have associated spin equal to $-1$, 
and the fourth has associated spin equal to $+1$, then 
\begin{displaymath}
H(\sigma)-H(\sigma^{i,0})= h
\end{displaymath}
\item\label{i:bcflip02}
if there exists a site $i$ such that $\sigma(i)=-1$ and 
there are at most two nearest neighbor of $i$ such that the 
associated spin in $\sigma$ is $-1$, then 
\begin{displaymath}
H(\sigma)-H(\sigma^{i,0})\ge h
\end{displaymath}
\item\label{i:bcflip03}
if there exists a site $i$ such that $\sigma(i)=0$,
there are at most two nearest neighbors of $i$ such that the 
associated spins are $0$, and the remaining nearest neighbors 
have associated spin equal to $+1$, then 
\begin{displaymath}
H(\sigma)-H(\sigma^{i,+1})\ge h
\end{displaymath}
\end{enumerate}
\end{lemma}

\medskip
Proof of Lemma~\ref{t:bcflip}.\
Items~\ref{i:bcflip01} and \ref{i:bcflip02}: 
let $n_+,n_-,n_0\ge0$ be the number of plus, minus, and zero 
spins, respectively, among the four nearest neighbors of the site $i$.
Note that $n_++n_-+n_0=4$.
From (\ref{ham-bc0}),
use also that here $\sigma(i)=-1$,
it follows immediately that 
\begin{displaymath}
H(\sigma)-H(\sigma^{i,0})
=
4n_++n_0+h-[n_-+n_+]
=
h+3n_+-n_-+n_0
\end{displaymath}
By using that $n_++n_0=4-n_-$ we get 
\begin{displaymath}
H(\sigma)-H(\sigma^{i,0})
=
h+4+2(n_+-n_-)
\end{displaymath}
Item~\ref{i:bcflip01} follows by the formula above 
with $n_-=3$ and $n_+=1$;
item~\ref{i:bcflip02} follows by the formula above 
by noticing that 
$n_-\le2$ and $n_+\ge0$. 

Item~\ref{i:bcflip03}:
let $n_+,n_0\ge0$ be the number of plus and zero 
spins, respectively, among the four nearest neighbors of the site $i$.
Note that, by hypothesis, $n_++n_0=4$.
From (\ref{ham-bc0}),
use also that here $\sigma(i)=0$,
it follows immediately that 
\begin{displaymath}
H(\sigma)-H(\sigma^{i,+1})
=
n_+-[n_0-h]
=
4-2n_0+h
\end{displaymath}
where we have used $n_+=4-n_0$.
The item finally follows by noticing that, by hypothesis, $n_0\le2$.
\qed

\medskip
Proof of item~\ref{i:mdep-bc01-02}
of Lemma~\ref{t:mdep-bc01}.
Let $\sigma\in\cc{X}\setminus\{\muno,\zero,\puno\}$, we have to prove 
that $V_\sigma<\Gamma_\rr{c}$; that is we have to find a configuration 
at energy smaller than $H(\sigma)$ and a path connecting $\sigma$ 
to such a configuration such that its height is smaller than 
$H(\sigma)+\Gamma_\rr{c}$.

We first assume that $\sigma$ is such that 
at least one spin is equal to minus one and
we distinguish several different sub--cases.

\smallskip
\par\noindent
Case $1$.\
There exists a plus--minus interface in $\sigma$, that is there exist 
$i,j\in\Lambda$ nearest neighbors such that 
$\sigma(i)=-1$ and $\sigma(j)=+1$.
From items~\ref{i:bcflip01} and \ref{i:bcflip02} in Lemma~\ref{t:bcflip}
we have that 
$H(\sigma)-H(\sigma^{i,0})\ge h$. 
It then follows 
$H(\sigma)>H(\sigma^{i,0})$ and, hence, $V_\sigma=0$.

\smallskip
\par\noindent
Case $2$.\
No plus--minus interface exist in $\sigma$ and there exist a 
site $i\in\Lambda$ such that $\sigma(i)=-1$ and at least two 
of its nearest neighbors are occupied by spin zero. 
From item~\ref{i:bcflip02} in Lemma~\ref{t:bcflip}
we have that 
$H(\sigma)-H(\sigma^{i,0})\ge h$. 
It then follows 
$H(\sigma)>H(\sigma^{i,0})$ and, hence, $V_\sigma=0$.

\smallskip
\par\noindent
Case $3$.\
No plus--minus interface exist in $\sigma$ and for any 
$i\in\Lambda$ such that $\sigma(i)=-1$ there is at most one 
nearest neighbor occupied by spin zero. 
Let $\gamma$ be the collection
of the unit segments of the dual of $\Lambda$ separating two 
sites with which are associated a minus and a zero spin; call 
those sites \definisco{adjacent} to $\gamma$.
This collection $\gamma$ is made of maximal connected components which 
are either rectangles or straight annuli winded around the torus. 
Note that by the characterization of $\sigma$ given in this case $3$
we have that each minus associated with a site adjacent to $\gamma$ 
has one single zero (the one associated to the neighbor adjacent to 
$\gamma$) among its nearest neighbors.

\smallskip
\par\noindent
Case $3.1$.\
Suppose that one of the connected component
of $\gamma$ is a straight annulus $\alpha$ winding around the torus. 
The annulus separates a stripe of zeros
by a stripe of minuses (both winding around the torus). 
Since each minus associated with a site adjacent to $\alpha$ 
has one single zero among its nearest neighbors, it follows 
that the minus stripe occupies at least two lattice rows or columns.
Then, 
let $\sigma'$ be the configuration obtained by flipping to zero 
all the minuses of $\sigma$ associated with sites adjacent to $\alpha$ and 
note that, by (\ref{ham-bc0}),  
\begin{displaymath}
H(\sigma')-H(\sigma)=-hL
\Rightarrow
H(\sigma')<H(\sigma)
\end{displaymath}
where we recall that $L$ is the side length of $\Lambda$. 
Moreover, consider the $L+1$ long 
path connecting $\sigma$ to $\sigma'$ constructed by 
flipping to zero first one of the minus spin associated with a site 
adjacent to $\alpha$ and then the remaining $L-1$ minuses so that 
at each step one minus with two zeros among its nearest neighbors 
is flipped. The height of this path is $H(\sigma)+2-h$. 
Since $\Gamma_\rr{c}>2-h$ we have that $V_\sigma<\Gamma_\rr{c}$.

\smallskip
\par\noindent
Case $3.2$.\
Suppose that none of the connected component
of $\gamma$ is a straight annulus winding around the torus
and assume that there exists a rectangular component  
such that one of its four sides, say $\rho$, has length 
$|\rho|$ larger or equal to $\ell_\rr{c}$.
Let $\sigma'$ be the configuration obtained by flipping to zero 
all the minuses of $\sigma$ associated with sites adjacent to $\rho$.
Since each minus associated with a site adjacent to $\rho$ 
has one single zero among its nearest neighbors,
by (\ref{ham-bc0}), we have that 
\begin{displaymath}
H(\sigma')-H(\sigma)
=2-h|\rho|
\le2-h\ell_\rr{c}
<2-h\frac{2}{h}
\Rightarrow
H(\sigma')<H(\sigma)
\end{displaymath}
where we have used that $\ell_\rr{c}>2/h$ (recall (\ref{lc})).
Moreover, consider the $|\rho|+1$ long 
path connecting $\sigma$ to $\sigma'$ constructed by 
flipping to zero first one of the minus spin associated with a site 
adjacent to $\rho$ and then the remaining $|\rho|-1$ minuses so that 
at each step one minus with two zeros among its nearest neighbors 
is flipped. The height of this path is $H(\sigma)+2-h$. 
Since $\Gamma_\rr{c}>2-h$ we have that $V_\sigma<\Gamma_\rr{c}$.

\smallskip
\par\noindent
Case $3.3$.\
Suppose that none of the connected component
of $\gamma$ is a straight annulus winding around the torus
and assume that there exists a rectangular component  
such that one of its four sides, say $\rho$, has length 
$|\rho|=1$. 
Let $\sigma'$ be the configuration obtained by flipping to minus
the zero of $\sigma$ associated with sites adjacent to $\rho$.
By considering the two possible situations, that is to say the 
$1\times1$ square and the 
$1\times\ell$ rectangle, 
by (\ref{ham-bc0})
we have that 
\begin{displaymath}
H(\sigma')-H(\sigma)
\le
\max\{-4+h,-2+h\}
=-2+h
<0
\Rightarrow
H(\sigma')<H(\sigma)
\end{displaymath}
where we have used that $h<1$.
Since $\sigma$ is transformed in $\sigma'$ in one step, we have that 
$V_\sigma=0<\Gamma_\rr{c}$.

\smallskip
\par\noindent
Case $3.4$.\
Suppose that none of the connected component
of $\gamma$ is a straight annulus winding around the torus
and assume that there exists a rectangular component  
such that one of its four sides, say $\rho$, has length 
$|\rho|=2$. 
Let $\sigma'$ be the configuration obtained by flipping to minus
the two zeros of $\sigma$ associated with sites adjacent to $\rho$.
By (\ref{ham-bc0})
we have that 
\begin{displaymath}
H(\sigma')-H(\sigma)
=
-2+2h
<0
\Rightarrow
H(\sigma')<H(\sigma)
\end{displaymath}
where we have used that $h<1$.
Moreover, consider the  
path long three connecting $\sigma$ to $\sigma'$ constructed by 
flipping to minus the two zero spins associated with the sites 
adjacent to $\rho$.
The height of this path is $H(\sigma)+h$. 
Since $\Gamma_\rr{c}>h$ we have that $V_\sigma<\Gamma_\rr{c}$.

\smallskip
\par\noindent
Case $3.5$.\
Suppose that none of the connected component
of $\gamma$ is a straight annulus winding around the torus
and assume that all the rectangular components have side 
lengths larger or equal to three and smaller or equal to $\ell_\rr{c}-1$.
Note that this case can be empty if $\ell_\rr{c}=3$. 
Consider one of this rectangular component and let $\rho$ be 
one of its four sides; note that 
$3\le|\rho|\le\ell_\rr{c}-1$.
For the sake of clearness assume that $\rho$ is vertical.

\smallskip
\par\noindent
Case $3.5.1$.\
Suppose that the column of sites at distance one from those occupied by zeros 
and adjacent to $\rho$ 
(in other less precise words the ``second" column inside the rectangle 
starting from $\rho$)
is occupied only by spin zero. 
Let $\sigma'$ be the configuration obtained by flipping to minus
the $|\rho|$ zeros of $\sigma$ associated with sites adjacent to $\rho$.
By (\ref{ham-bc0})
we have that 
\begin{displaymath}
H(\sigma')-H(\sigma)
=
-2+h|\rho|
\le
-2+h(\ell_\rr{c}-1)
<
-2+h\frac{2}{h}
<0
\Rightarrow
H(\sigma')<H(\sigma)
\end{displaymath}
where we have used that $|\rho|\le\ell_\rr{c}-1$.
Moreover, consider the $|\rho|+1$ long 
path connecting $\sigma$ to $\sigma'$ constructed by 
flipping to minus, one after the others, 
the zero spins associated with the sites 
adjacent to $\rho$ and by flipping, in the first 
$|\rho|-1$ steps, one of the two zeros with two minuses among its four 
nearest neighbors; in the final step the remaining spin zero is 
finally flipped to minus.
The height of this path is $H(\sigma)+h(|\rho|-1)$. 
Since $\Gamma_\rr{c}>h(\ell_\rr{c}-2)$ we have that $V_\sigma<\Gamma_\rr{c}$.

\smallskip
\par\noindent
Case $3.5.2$.\
Suppose that the column of sites at distance one from those occupied by zeros 
and adjacent to $\rho$ 
(in other less precise words the ``second" column inside the rectangle 
starting from $\rho$)
is occupied at least by one spin plus. 
Let $\sigma'$ be the configuration obtained by flipping to zero
the pluses of $\sigma$ associated with sites at distance one from those  
adjacent to $\rho$ and occupied by zeros.

\smallskip
\par\noindent
Case $3.5.2.1$.\
Suppose $|\rho|-2=1$.
The configuration $\sigma$ is transformed into $\sigma'$ by flipping to 
zero a plus spin having at least three zeros among its 
nearest neighbors and no minus. 
It then follows that 
\begin{displaymath}
H(\sigma')-H(\sigma)
\le
\max\{-4+h,-2+h\}
=-2+h
<0
\Rightarrow
H(\sigma')<H(\sigma)
\end{displaymath}
where we have used that $h<1$.
Since $\sigma$ is transformed into $\sigma'$ in one step we have that 
$V_\sigma<0<\Gamma_\rr{c}$. 

\smallskip
\par\noindent
Case $3.5.2.2$.\
Suppose $|\rho|-2\ge2$.
The configuration $\sigma$ is transformed into $\sigma'$ by flipping to 
zero at most $|\rho|-3$ plus spin having two zeros and two pluses as nearest 
neighbors and 
at least one plus spin having no minus and at most one plus 
among its nearest neighbors. 
It then follows that 
\begin{displaymath}
H(\sigma')-H(\sigma)
\le
\max\{-4+h,-2+h\}+(|\rho|-3)h
\le
-2+(|\rho|-2)h
\end{displaymath}
Hence, recalling that $|\rho|\le\ell_\rr{c}-1$, we have 
\begin{displaymath}
H(\sigma')-H(\sigma)
\le
-2+(\ell_\rr{c}-3)h
<
-2+\frac{2}{h}\,h-2h
<-2h
\Rightarrow
H(\sigma')<H(\sigma)
\end{displaymath}
Moreover, consider the  
path connecting $\sigma$ to $\sigma'$ constructed by 
flipping to zero, one after the others, 
the plus spins with two neighboring pluses and then those 
with at most one neighboring plus. 
The height of this path is smaller that 
$H(\sigma)+h(|\rho|-3)$. 
Since $\Gamma_\rr{c}>h(\ell_\rr{c}-4)$ we have that $V_\sigma<\Gamma_\rr{c}$.

\medskip
In order to complete the proof, 
we finally have to consider those configurations 
$\sigma\in\cc{X}\setminus\{\muno,\zero,\puno\}$ such that 
none of the spins is minus. 
The proof can be achieved with arguments very similar to those developed 
above; we just sketch the idea.
First of all we consider the collection
of the unit segments of the dual of $\Lambda$ separating two 
sites with which are associated a zero and a plus spin.
From item~\ref{i:bcflip03} in Lemma~\ref{t:bcflip} it follows that
this collection is made of maximal connected components which 
are either rectangles or straight annuli winded around the torus. 
We then have to consider all the cases analogous to those 
taken into account above.
\qed

\medskip
Proof of Theorem~\ref{t:meta-bc01}.
The theorem follows from Lemma~\ref{t:mdep-bc01} 
and Theorem~\ref{t:schemamnos}.
\qed

\medskip
Proof of Theorem~\ref{t:meta-bc02}.
The theorem follows from Theorem~\ref{t:meta-bc01} 
above
and
\cite[Theorems~4.1 and 4.9]{MNOS}.
\qed

\subsection{Escape mechanism}
\label{s:pnuclea}
\par\noindent
In this subsection we prove Lemma~\ref{t:mdep-bc02} and 
Theorem~\ref{t:meta-bc03}
allowing the identification of the escape mechanism as the nucleation 
of the critical droplet. 

\medskip
Proof of Lemma~\ref{t:mdep-bc02}.
Item~\ref{i:mdep-bc02-01}:
the proof of this item is divided into two parts. 
In the first part we characterize the set $F(\vbim)$ 
of the minima of the energy 
of the set $\vbim\subset \cc{X}$ defined at the beginning 
of Subsection~\ref{s:pmetlat}.
In the second part we show 
that any path connecting $\muno$ to $\puno$ must necessarily 
pass through $\pcri$.

\begin{figure}[t]
 \begin{picture}(100,150)(-50,25)
 \setlength{\unitlength}{0.08cm}
 \thinlines
 \qbezier[50](0,15)(22.5,15)(45,15)
 \qbezier[50](0,60)(22.5,60)(45,60)
 \qbezier[50](0,15)(0,37.5)(0,60)
 \put(0,25){\line(0,1){15}}
 \put(0,40){\line(1,0){10}}
 \put(10,40){\line(0,1){5}}
 \put(10,45){\line(1,0){15}}
 \put(25,45){\line(0,1){15}}
 \put(25,60){\line(1,0){20}}
 \put(45,60){\line(0,-1){45}}
 \put(45,15){\line(-1,0){15}}
 \put(30,15){\line(0,1){5}}
 \put(30,20){\line(-1,0){15}}
 \put(15,20){\line(0,1){5}}
 \put(15,25){\line(-1,0){15}}
 \put(60,15){\line(1,0){40}}
 \put(60,15){\line(0,1){45}}
 \put(60,60){\line(1,0){5}}
 \put(70,60){\line(1,0){30}}
 \put(70,60){\line(0,1){5}}
 \put(65,65){\line(1,0){5}}
 \put(65,60){\line(0,1){5}}
 \put(100,15){\line(0,1){45}}
 \qbezier[50](60,65)(80,65)(100,65)
 \qbezier[7](60,60)(60,62.5)(60,65)
 \qbezier[7](100,60)(100,62.5)(100,65)
 \put(115,15){\line(1,0){40}}
 \put(115,15){\line(0,1){45}}
 \put(115,60){\line(1,0){40}}
 \put(155,15){\line(0,1){10}}
 \put(155,60){\line(0,-1){30}}
 \put(155,25){\line(1,0){5}}
 \put(155,30){\line(1,0){5}}
 \put(160,25){\line(0,1){5}}
 \qbezier[7](155,15)(157.5,15)(160,15)
 \qbezier[7](155,60)(157.5,60)(160,60)
 \qbezier[50](160,15)(160,37.5)(160,60)
 \end{picture}
 \caption{Typical configuration of $F_1(\vbim)$ (left) 
          and $F_2(\vbim)$ (center and right).
          Zeros are associated with the sites inside the 
          solid lines and minuses outside.}
 \label{f:fig04}
\end{figure}
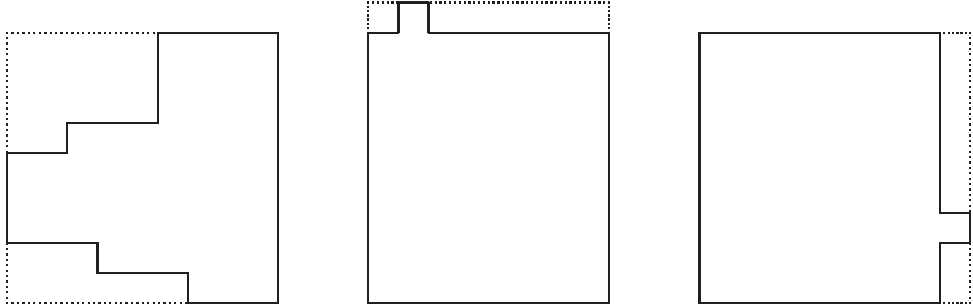

\medskip
First part of the proof.\
From items~\ref{i:vbim01} and \ref{i:vbim02} of Lemma~\ref{t:vbim}
it follows that 
for any $\sigma\in F(\vbim)$
the polyomino $C(\sigma)$ associated with the zero spins does not
wind around the torus. We can then apply 
item~\ref{i:polyomini0-03} in
Lemma~\ref{t:polyomini0} and deduce that 
$C(\sigma)$ is convex. 
Thus,
we partition the set $F(\vbim)$ into two disjoint subsets
$F_1(\vbim)$ and $F_2(\vbim)$ defined as follows (see, also, 
figure~\ref{f:fig04}):
the set $F_1(\vbim)$  
is the set of configurations $\sigma\in F(\vbim)$ such 
that the boundary of the polyomino $C(\sigma)$ intersects 
each side of the boundary of its smallest surrounding rectangle
on a set of the dual lattice $\bb{Z}^2+(1/2,1/2)$ made by 
at least two pairwise consecutive unit segments. 
The set $F_2(\vbim)$  
is the set of configurations $\sigma\in F(\vbim)$ such 
that the boundary of the polyomino $C(\sigma)$ intersects 
at least one of the four 
sides of the boundary of its smallest surrounding rectangle
on a set made of a single unit segments. 
Note that, since $C(\sigma)$ is convex, there cannot exist 
multiple unit protuberances intersecting the same side 
of the smallest surrounding rectangle.  

In figure~\ref{f:fig04} the two configurations in $F_2(\vbim)$ 
have been represented by a rectangle of zeros with a unit 
zero protuberance placed either on the longest (picture on the right)
or on the shortest side (center picture). 
In principle other situations should be taken into 
account, indeed the boundary of the 
polyomino associated with the zero component of the 
considered configuration
could intersect the other three sides of the boundary of 
its smallest surrounding rectangle in proper subsets of the side 
itself. In other words the rectangle on which the unit protuberance 
is placed could be not fully occupied by zeros. 

We prove, now, that the two situations depicted in the picture indeed cover 
all the possible cases. 
Consider $\sigma\in F_2(\vbim)$.
Let
$\ell_\rr{c}+k$ and $\ell_\rr{c}+s$ be the side lengths of 
the smallest rectangle surrounding the polyomino $C(\sigma)$ 
associated with $\sigma$ with $k,s\in\bb{Z}$. 
Assume, also, that a single protuberance $U$ is placed on the side of 
length $\ell_\rr{c}+s$. 
Let $R(\sigma)$ be the smallest rectangle surrounding the 
polyomino obtained by removing the unit cell $U$ 
from $C(\sigma)$;
note that $R(\sigma)$ is a rectangle with side lengths 
$\ell_\rr{c}+s$ and $\ell_\rr{c}+k-1$.
Since the polyomino $C(\sigma)$ is convex, we have that its 
perimeter is equal to that of the smallest surrounding 
rectangle, namely, 
\begin{displaymath}
\rr{Perimeter}(C(\sigma))
=2(\ell_\rr{c}+k)+2(\ell_\rr{c}+s)=4\ell_\rr{c}+2(k+s)
\end{displaymath}
Moreover, from item~\ref{i:vbim04} in Lemma~\ref{t:vbim} it follows that 
$\rr{Perimeter}(C(\sigma))=4\ell_\rr{c}$
and, hence, $k+s=0$.

Note, now, that the area of the rectangle $R(\sigma)$ is  
\begin{displaymath}
(\ell_\rr{c}+s)(\ell_\rr{c}+k-1)
=
\ell_\rr{c}(\ell_\rr{c}-1)+(k+s)\ell_\rr{c}+ks-s
=
\ell_\rr{c}(\ell_\rr{c}-1)-s^2-s
\end{displaymath}
where we have used $k+s=0$.
Since the area of the polyomino $C(\sigma)$ is $\ell_\rr{c}(\ell_\rr{c}-1)+1$, 
the number of zeros inside $R(\sigma)$ is $\ell_\rr{c}(\ell_\rr{c}-1)$.
Hence the area of $R(\sigma)$ must be larger or equal to 
$\ell_\rr{c}(\ell_\rr{c}-1)$; it then follows that 
\begin{displaymath}
\ell_\rr{c}(\ell_\rr{c}-1)-s^2-s
\ge
\ell_\rr{c}(\ell_\rr{c}-1)
\Rightarrow
-s^2-s\ge0
\end{displaymath}
Since $s$ is an integer, the above inequality is satisfied only for 
$s=0$ and $s=-1$; note that in both cases the inequality is indeed 
an equality.
We can then give a full characterization of the set $F_2(\vbim)$.
Indeed we can write 
\begin{displaymath}
F_2(\vbim)=F_{2,a}(\vbim)\cup F_{2,b}(\vbim)
\end{displaymath}
with 
$F_{2,a}(\vbim)=\pcri$
and 
$F_{2,b}(\vbim)$ the set of configurations of $F(\vbim)$ 
such that the zeros occupy a rectangle of side lengths 
$\ell_\rr{c}-1$ and $\ell_\rr{c}$ and a unit protuberance 
placed on one of the two shortest sides.

\medskip
Second part of the proof.\
We prove, now, that any path connecting $\muno$ to $\puno$ necessarily
pass through $\pcri$. The proof is organized in several steps. 

Step 0:
for any path $\omega\in\Omega(\muno,\puno)$
there exists a positive integer $i$ such that 
$\omega_i\in\vbim$.
See the proof (lower bound) 
of item~\ref{i:mdep-bc01-01} of Lemma~\ref{t:mdep-bc01}.

Step 1:
for any path $\omega\in\Omega(\muno,\puno)$ such that 
$\Phi_\omega-H(\muno)=\Gamma_\rr{c}$
there exists a positive integer $i$ such that 
$\omega_i\in F(\vbim)$.
This property follows from step~0 and 
item~\ref{i:vbim04} in Lemma~\ref{t:vbim}.

Step 2:
for any path $\omega\in\Omega(\muno,\puno)$
such that $\Phi_\omega-H(\muno)=\Gamma_\rr{c}$
we denote by $f(\omega)$ 
the set of positive integers such that for any $i\in f(\omega)$
one has that
$\omega_i\in F(\vbim)$
and $\omega_{i-1}$ is
made of $|\Lambda|-\ell_\rr{c}(\ell_\rr{c}-1)$ minus spins and 
$\ell_\rr{c}(\ell_\rr{c}-1)$ zeros. 
From step~1 above it follows that $f(\omega)$ is not empty.

Step 3:
for any path $\omega\in\Omega(\muno,\puno)$
such that $\Phi_\omega-H(\muno)=\Gamma_\rr{c}$ 
one has that 
$\omega_i\in F_2(\vbim)$ for any $i\in f(\omega)$. 
Indeed, assume by absurdity that
that there exists $i\in f(\omega)$ 
such that $\omega_i\in F_1(\vbim)$.
By definition of $f(\omega)$ the configuration $\omega_i$ is 
transformed into $\omega_{i-1}$ by flipping to minus one of the 
zeros in $\omega_i$. 
From the definition 
of $F_1(\vbim)$ it follows that all the zero spins in $\omega_i$ 
have at least two zeros among their nearest neighbors. 
Then by (\ref{ham-bc0}) it follows easily that 
$H(\omega_{i-1})>H(\omega_i)$.
Hence, 
\begin{displaymath}
\Phi_\omega
\ge
H(\omega_{i-1})
>
H(\omega_i)
=
H(F(\vbim))
=
H(\muno)+\Gamma_\rr{c}
\end{displaymath}
which is an absurd. We can then conclude that $\omega_i\in F_2(\vbim)$
for any $i\in f(\omega)$. 

Step 4:
for any path $\omega\in\Omega(\muno,\puno)$
such that $\Phi_\omega-H(\muno)=\Gamma_\rr{c}$,
if $i$ is a positive integer such that $\omega_i\in F(\vbim)$ 
then
$\omega_{i-1},\omega_{i+1}\not\in \vbim$.
Indeed, assume by absurdity that $\omega_{i+1}\in \vbim$. 
Then $\omega_{i+1}$ is necessarily obtained by $\omega_i$ by flipping 
to plus a zero spin, otherwise the number of minus spins 
would be changed (recall that all the configurations in $\vbim$ have 
the same number of minuses).
Since any zero spin of $\omega_i$ has no plus among its nearest
neighbors, from (\ref{ham-bc0}) it follows that 
$H(\omega_{i+1})>H(\omega_i)$. Hence, 
\begin{displaymath}
\Phi_\omega
\ge
H(\omega_{i+1})
>
H(\omega_i)
=
H(F(\vbim))
=
H(\muno)+\Gamma_\rr{c}
\end{displaymath}
which is an absurd. We can then conclude that $\omega_{i+i}\not\in\vbim$.
Similarly we also prove that $\omega_{i-i}\not\in\vbim$.

\begin{figure}[t]
 \begin{picture}(400,210)(-60,-15)
 \setlength{\unitlength}{0.03cm}
 \thinlines
 \qbezier(0,100)(0,200)(200,200)
 \qbezier(0,100)(0,0)(200,0)
 \qbezier(200,200)(400,200)(400,100)
 \qbezier(200,0)(400,0)(400,100)
 \qbezier(200,0)(100,100)(200,200)
 \qbezier(45,55)(55,220)(182,180)
 \qbezier(182,180)(100,180)(168,160)
 \qbezier(168,160)(100,170)(158,140)
 \qbezier(158,140)(200,120)(155,130)
 \qbezier(155,130)(200,110)(152,120)
 \qbezier(152,120)(200,100)(151,110)
 \qbezier(151,110)(100,120)(150,100)
 \qbezier(150,100)(100,100)(152,80)
 \qbezier(152,80)(200,60)(158,60)
 \qbezier(158,60)(200,50)(163,50)
 \qbezier(163,50)(200,40)(175,30)
 \qbezier(175,30)(200,20)(195,5)
 \qbezier(195,5)(300,140)(354,50)
 \put(340,5){$\cc{X}$}
 \put(20,115){$\cc{X}_{\rr{d},>}$}
 \put(350,115){$\cc{X}_{\rr{d},<}$}
 \put(195,185){$\cc{X}_{\rr{d}}$}
 \put(40,40){$\muno$}
 \put(350,40){$\puno$}
 \put(152,80){\circle*{5}}
 \put(155,83){${\scriptstyle \omega_j}$}
 \end{picture}
 \caption{Paths constructed in the second part of the proof 
          of Lemma~\ref{t:mdep-bc02}.}
 \label{f:fig05}
\end{figure}

Step 5: 
consider a path $\omega\in\Omega(\muno,\puno)$
such that $\Phi_\omega-H(\muno)=\Gamma_\rr{c}$
and assume, by absurdity, that it does not pass through $\pcri$, 
that is to say $\omega_i\not\in\pcri$ for any integer $i$. 

From step~4 (see figure~\ref{f:fig05})
it follows that there exists $n$ not consecutive 
integers $i_1<i_2<\cdots<i_{n-1}<i_n$ such that 
$\omega_{i_k}\in F(\vbim)$, for $k=1,\dots,n$.
If $n\ge2$,
each sub--path 
$(\omega_{i_k+1},\dots,\omega_{i_{k+1}-1})$, for $k=1,\dots,n-1$,
belongs either 
to the subset $\vbiminore$ of $\cc{X}$ made of all the configurations 
with number of minus spins smaller or equal to 
$|\Lambda|-[\ell_\rr{c}(\ell_\rr{c}-1)+1]-1$
or 
to the subset $\vbimaggiore$ of $\cc{X}$ made of all the configurations 
with number of minus spins larger or equal to 
$|\Lambda|-[\ell_\rr{c}(\ell_\rr{c}-1)+1]+1$.

Let $j$ be the smallest integer in $\{1,\dots,n\}$ such that 
the path $(\omega_{i_j},\omega_{{i_j}+1},\dots,\puno)$ 
does not visit $\vbimaggiore$. Since 
$\omega_{{i_j}-1}\in\vbimaggiore$, from step~3
and from the absurd hypothesis 
it follows that $\omega_{i_j}\in F_{2,b}(\vbim)$. 

Since $H(\omega_{i_j})=H(\muno)+\Gamma_\rr{c}$, 
starting from $\omega_{i_j}$ the path must necessarily decrease the 
energy. 
Since it is not possible 
to flip to minus the unit zero protuberance (otherwise the path
would enter $\vbimaggiore$), the only move that decreases the 
energy is flipping to zero a minus with two zeros among 
its nearest neighbors. So that 
$H(\omega_{{i_j}+1})=H(\muno)+\Gamma_\rr{c}-h$.

Starting from $\omega_{{i_j}+1}$ only moves which increase
the energy of at most $h$ are allowed. So that the sole possible 
moves are: flipping to zero a minus with two zeros among its 
nearest neighbors or flipping to minus a zero with two minuses 
among its nearest neighbors. 

With these types of moves only configurations $\sigma$  such that the polyomino 
$C(\sigma)$ associated with the zero spin component is convex 
and its smallest surrounding rectangle has side lengths 
$\ell_\rr{c}+1$ and $\ell_\rr{c}-1$. 
In this set the smallest energy is that of the configuration 
in which $C(\sigma)$ is precisely the rectangle with side lengths 
$\ell_\rr{c}+1$ and $\ell_\rr{c}-1$. 

Besides the already mentioned moves, the one to which 
competes the smallest energy increase $2-h$ is flipping to zero 
a minus with one zero and three minuses among its nearest 
neighbors. Thus, we have that 
\begin{displaymath}
\Phi_\omega-H(\muno)
\ge 
2(\ell_\rr{c}+1)+2(\ell_\rr{c}-1)-
h(\ell_\rr{c}+1)(\ell_\rr{c}-1)+2-h
\end{displaymath}
Hence
\begin{displaymath}
\Phi_\omega-H(\muno)
\ge 
4\ell_\rr{c}-
h[\ell_\rr{c}(\ell_\rr{c}-1)+1]
-h(\ell_\rr{c}-1)+2
>\Gamma_\rr{c}
\end{displaymath}
where, in the last inequality, we have used (\ref{gammac0}), 
(\ref{gammac}), and the bound
\begin{displaymath}
2-h(\ell_\rr{c}-1)
>
2-h\frac{2}{h}
>0
\end{displaymath}
where we recalled the upper bound (\ref{lcp}).

We finally got an absurd. We then have that the path $\omega$ has to 
visit $\pcri$. 
This completes the proof of 
item~\ref{i:mdep-bc02-01}
of the lemma.

Item~\ref{i:mdep-bc02-02}:
the proof 
can be achieved, ``mutatis mutandis," with the same arguments 
used in the proof of item~\ref{i:mdep-bc02-01}.
We do not enter into the 
details, we just remark that the manifold $\vbim$ must be 
replaced by $\vbizero$ defined as 
the set of configurations in which all the spins are zeros excepted 
for $\ell_\rr{c}(\ell_\rr{c}-1)+1$ which are pluses.
\qed

\medskip
Proof of Theorem~\ref{t:meta-bc03}.
In the proof of this theorem we use results in \cite{MNOS}, 
in particular the definition of gate on page~603 and 
the Theorem~5.4.
The Lemma~\ref{t:mdep-bc02} in Section~\ref{s:bcmodel} above
implies that $\pcri$ is a gate for the 
pair of configurations $\muno$ and $\puno$
and 
$\qcri$ is a gate for the 
pair of configurations $\zero$ and $\puno$.
Then the theorem follows from Theorem~5.4 in \cite{MNOS}.
\qed

\medskip
Proof of Theorem~\ref{t:mdep-bc04}.
As stated in item~\ref{i:mdep-bc0.5-01.5}
of Lemma~\ref{t:mdep-bc0.5} (see also figure~\ref{f:fig03}), 
there exists a path joining $\muno$ to $\puno$ 
passing through any state in $\pcri$ and attaining its maximal height only in 
this point. This implies that any subset of $\cc{X}$ not containing 
$\pcri$ is not a gate, in particular it is not minimal. 

The second statement of the theorem can be prove similarly.
\qed


\end{document}